\documentclass[aps,prl,twocolumn,showpacs,groupedaddress,superscriptaddress,footinbib,longbibliography]{revtex4-2}
\usepackage{mathrsfs}
\usepackage{natbib,hyperref}
\setcitestyle{square,sort&compress,comma,numbers}
\hypersetup{colorlinks=true, citecolor=blue, urlcolor=blue, linkcolor=blue}
\usepackage[english]{babel}
\usepackage[utf8]{inputenc}
\usepackage{graphicx}
\usepackage[caption=false]{subfig}
\usepackage{amsmath}
\usepackage{amsfonts}
\usepackage{amssymb}
\usepackage{color,soul}
\setulcolor{red}
\usepackage{easyReview}
\usepackage{xcolor}
\usepackage{tikz}
\usepackage{pgfplots}
\usepackage{ulem}
\usepgfplotslibrary{colormaps}
\pgfplotsset{
    colormap={mycolormap}{
        rgb255=(59,76,192)
        rgb255=(255,255,255)
        rgb255=(180,4,38)
    }}

\renewcommand{\vec}[1]{\mathbf{#1}}

\begin{document}

\title{Vorticity-induced surfing and trapping in porous media}
\author{Pallabi Das}
\affiliation{Max Planck Institute for the Physics of Complex Systems, N\"ohnitzer Stra{\ss}e 38,
01187 Dresden, Germany}
\author{Mirko Residori}
\affiliation{Max Planck Institute for the Physics of Complex Systems, N\"ohnitzer Stra{\ss}e 38,
01187 Dresden, Germany}
\author{Axel Voigt}
\affiliation{Institute of Scientific Computing, Technische Universit\"at Dresden, 01062 Dresden, Germany}
\affiliation{Center for Systems Biology Dresden, Pfotenhauerstr. 108, 01307 Dresden, Germany}
\affiliation{Cluster of Excellence, Physics of Life, TU Dresden, Arnoldstr. 18, 01307 Dresden, Germany}
\author{Suvendu Mandal}
\affiliation{Technische Universit\"at Darmstadt, Karolinenplatz 5, 64289 Darmstadt, Germany}
\author{Christina Kurzthaler}
\email{ckurzthaler@pks-mpg.de}
\affiliation{Max Planck Institute for the Physics of Complex Systems, N\"ohnitzer Stra{\ss}e 38,
01187 Dresden, Germany}
\affiliation{Center for Systems Biology Dresden, Pfotenhauerstr. 108, 01307 Dresden, Germany}
\affiliation{Cluster of Excellence, Physics of Life, TU Dresden, Arnoldstr. 18, 01307 Dresden, Germany}

\begin{abstract} 
Microorganisms often encounter strong confinement and complex hydrodynamic flows while navigating their habitats. Combining finite-element methods and stochastic simulations, we study the interplay of active transport and heterogeneous flows in dense porous channels. We find that swimming always slows down the traversal of agents across the channel, giving rise to robust power-law tails of their exit-time distributions. These exit-time distributions collapse onto a universal master curve with a scaling exponent of $\approx 3/2$ across a wide range of packing fractions and motility parameters, which can be rationalized by a scaling relation. We further identify a new motility pattern where agents alternate between {\it surfing} along fast streams and extended {\it trapping} phases, the latter determining the power-law exponent. Unexpectedly, trapping occurs in the flow backbone itself -- not only at obstacle boundaries -- due to vorticity-induced reorientation in the highly-heterogeneous fluid environment. These findings provide a fundamentally new active transport mechanism with direct implications for biofilm clogging and the design of novel microrobots capable of operating in heterogeneous media.  
\end{abstract} 

\maketitle

Microorganisms are central to a wide range of biological processes, ranging from crop growth~\cite{Tecon:2017,Voigtlaender:2023},  bacterial infections~\cite{Croinin:2012, Kirsch:2019, Otte:2021}, and fertilization~\cite{Suarez:2006}, to biofilm formation and community ecology~\cite{Hartmann:2019, Hallatschek:2023}, to the carbon and nutrient cycle in the ocean~\cite{Seymour:2017, Nguyen:2021}. Beyond their ecological importance, they also serve as ideal model systems for the design of synthetic micro- and nanorobots that may deliver drugs to specific targets~\cite{Erkoc:2019}, penetrate the porous structure of tumors~\cite{Sedighi:2019}, and induce degradation of contaminants~\cite{Gao:2014, Adadevoh:2016, Li:2017, Carvalho:2018}. A prerequisite for many of these biological functions and applications is the ability of these agents to self-propel and respond to chemical and mechanical cues~\cite{Kurzthaler:2023}. While they can thereby overcome the constraints of diffusion, they often need to operate in complex environments characterized by geometric disorder~\cite{Kumar:2022, Spagnolie:2023, Jin:2024} and heterogeneous fluid flows~\cite{Guasto:2012, Wheeler:2019}, providing grand challenges for their efficient navigation.

Much of our current understanding comes from studies in channel-like geometries, where the interplay between swimming and shear has been shown to generate unusual behaviors. These include the reorientation of bacteria by the flow vorticity, leading to rheotaxis, i.e. motion along/against fluid flows~\cite{Mathijssen:2019, Peng:2020, Ezhilan:2015,Marcos:2012, Lagoin:2025}. A manifestation of this reorientation is the shape-dependent entrapment of active agents in regions of low or high fluid shear~\cite{Rusconi:2014, Barry:2015,Vennamneni:2020}. Furthermore, microswimmers have been observed -- both experimentally~\cite{Mino:2018, Secchi:2020} and theoretically~\cite{Lee:2021} -- to accumulate behind spherical obstacles, form boundary layers around them~\cite{Yan:2015}, and exhibit curly swimming trajectories~\cite{Dentz:2022}. Counter-intuitively, bacteria can accumulate after channel-wall constrictions in the direction of the flow~\cite{Altshuler:2013} or, in the presence of magnetic fields, display orbital swimming motion against the flow~\cite{waisbord2021fluidic}, highlighting interesting features resulting from the interplay of surface structure, flow, and activity. 

In contrast, far less is known about swimming in disordered environments, especially under flow. In the absence of flow, porous materials crucially modify the bacterial run-and-tumble dynamics to a hop-and-trap pattern, characterized by intermittent trapping phases in the corners of the pore space and hopping through the pores~\cite{Bhattacharjee:2019}. Theoretical work has further predicted optimal tumbling strategies for spreading through porous media~\cite{Bertrand:2018, Licata:2016, Volpe:2017, Kurzthaler:2021} and addressed the impact of external forces~\cite{Reichhardt:2014}. Yet, how external fluid flow, characterized by few fast streams and large stagnant zones~\cite{Residori:2025}, shapes microswimmer motion in dense porous media remains largely unexplored. 

Here, we address this gap by investigating how activity, hydrodynamic couplings, and geometric disorder jointly dictate transport across porous channels under flow. We show that increasing activity counter-intuitively slows down transport, reflected in robust power-law tails of the exit-time distributions with a universal exponent of $\approx 3/2$ for diverse packing fractions and motility parameters. Strikingly, vorticity promotes accumulation of active agents in the flow backbone, giving rise to prolonged trapping phases and a `surf-and-trap' motility pattern. This mechanism provides a physical basis for the onset of biofilm clogging in porous channels and biomedical devices~\cite{Aufrecht:2019, Kurz:2022, Kurz:2023}, while offering new design principles for microrobots functional in complex aqueous environments.

\begin{figure*}[tp]
\includegraphics[width=0.8\linewidth]{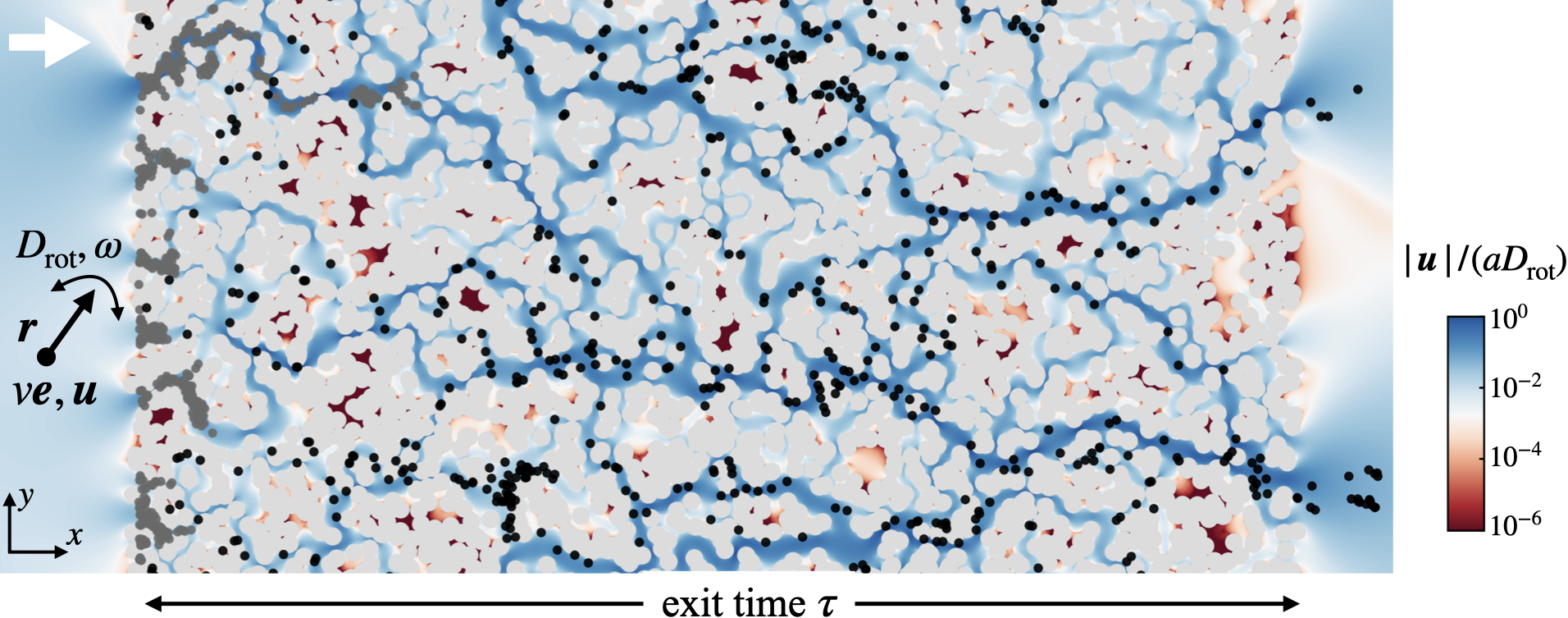}
    \caption{ \label{fig:method} {\bf Model set-up.} The active agent self-propels along its instantaneous orientation $\boldsymbol{e}$ at velocity $v$ and rotates due to rotational Brownian motion with diffusivity $D_{\rm rot}$. It is also advected and rotated by the velocity and vorticity fields, $\boldsymbol{u}$ and $\omega$, respectively. Simulation snapshots of swimmer positions at two consecutive times, $t_1=0.1/D_{\rm rot}$ and $t_2= 1.5/D_{\rm rot}$, are depicted by dark gray and black dots, respectively. The gray discs are the obstacles and the externally-applied flow is shown in the background. The color bar corresponds to the magnitude of the velocity field $|\boldsymbol{u}|$ and the white arrow indicates the direction of the applied flow. For illustration purpose we only show a slice of the porous channel. Here, the packing fraction and P{\'e}clet numbers are $\phi = 0.51$,  $\mathrm{Pe}^f = 40$, and $\mathrm{Pe}^s = 1$, respectively, and $a$ denotes the obstacle radius.}
\end{figure*}

\paragraph{Model.--} We model active agents as active Brownian particles (ABPs), which self-propel at velocity $v$ along their instantaneous orientation $\vec{e}$ and are subject to translational and rotational Brownian motion with diffusivities $D$ and $D_{\rm rot}$, respectively~\cite{Howse:2007,Romanczuk:2012,Kurzthaler:2016}. In addition, they move in a two-dimensional porous medium, modeled as randomly overlapping discs of radii $a$, in the presence of an externally-applied flow (Fig.~\ref{fig:method}). We assume that the ABPs do not perturb the fluid flow and, according to Fax\'{e}n's law~\cite{leal2007advanced}, are advected and rotated with the spatially-varying velocity and vorticity fields, $\boldsymbol{u}$ and~$\omega$, respectively (see Materials and Methods). A repulsive Weeks-Chandler-Andersen  potential is used to model swimmer-obstacle interactions~\cite{Zeitz:2017}. In this framework, two P{\'e}clet numbers emerge: the swim  P{\'e}clet number Pe$^s=v/(aD_{\rm rot})$ (also referred to as `activity') and the flow P\'{e}clet number ${\rm Pe}^f = \langle|\boldsymbol{u}|\rangle/(aD_{\rm rot})$. They characterize the relative importance of self-propulsion, respectively, average flow velocity to rotational diffusion. In addition, the effect of confinement is captured by the packing fraction $\phi=  1-{\rm exp}(-N\pi a^2/L^2)$ with length of the porous channel $L$ and number of obstacles~$N$~\cite{torquato_random_2002}. 

We quantify the transport of active agents in terms of their exit times $\tau$, i.e. the time it takes to traverse the porous channel in the direction of the applied flow. The ABPs start to swim from a random position and orientation in an obstacle-free region at the channel inlet. Typical particle positions are depicted in Fig.~\ref{fig:method}: While at short times  agents follow the same main flow path, at longer times their trajectories diversify due to the interplay of self-propulsion, advection, vorticity, and noise.

\begin{figure*}[tp]
\centering
\includegraphics[width=1.0\linewidth]{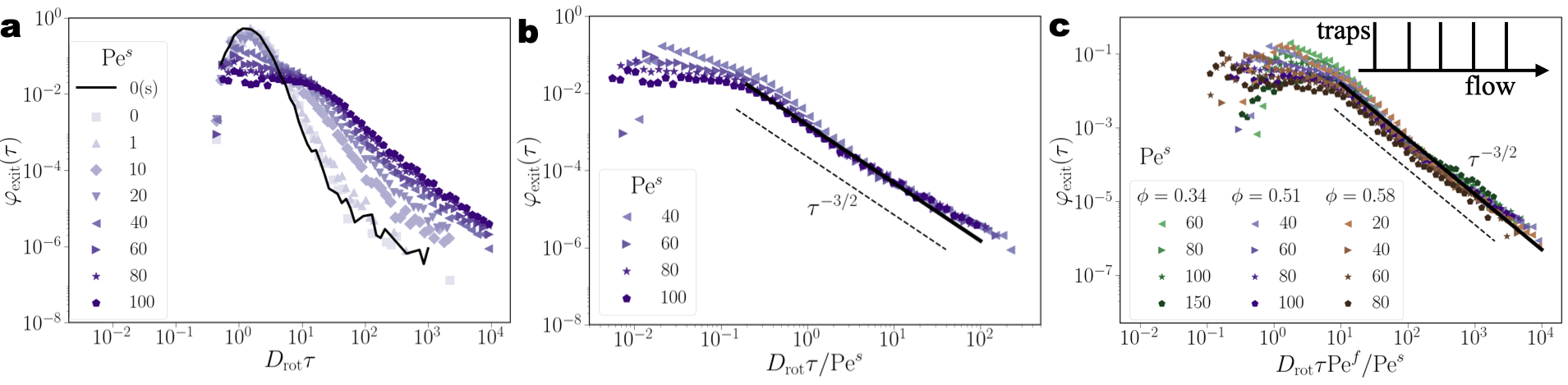}
\caption{ \label{fig:exittime} {\bf Power-law behavior of the exit-time distributions of active particles. a.}~Exit-time distribution $\varphi_{\rm exit}(\tau)$ for different swimming P{\'e}clet numbers. Here, the packing fraction and flow strength are $\phi=0.51$ and ${\mathrm{Pe}}^f=40$, respectively. The solid line represents the exit-time distribution of the streamlines of the fluid flow. {\bf b.}~Data collapse for $\mathrm{Pe}^s \gtrsim \mathrm{Pe}^f$ by rescaling the $x-$axis with ${\rm Pe}^s$. {\bf c.}~Data collapse for various packing fractions $\phi=0.34, 0.51, 0.58$ with their respective flow P{\'e}clet number $\mathrm{Pe}^f = 70,40,20$ and varying higher activity~${\rm Pe}^s$. The $x-$axis is rescaled by ${\rm Pe}^f/{\rm {Pe}^s}$. ({\it Inset})~Sketch of the proposed `comb potential' analog. The black lines in {\bf b}, {\bf c} indicate the power-law behavior at long times~$\propto \tau^{-3/2}$. The dashed black line is added as guide to the eye.  We denote by $D_{\rm rot}$ the rotational diffusivity.}
\end{figure*}

\begin{figure}[tp]
\centering
\includegraphics[width=1.0\linewidth]{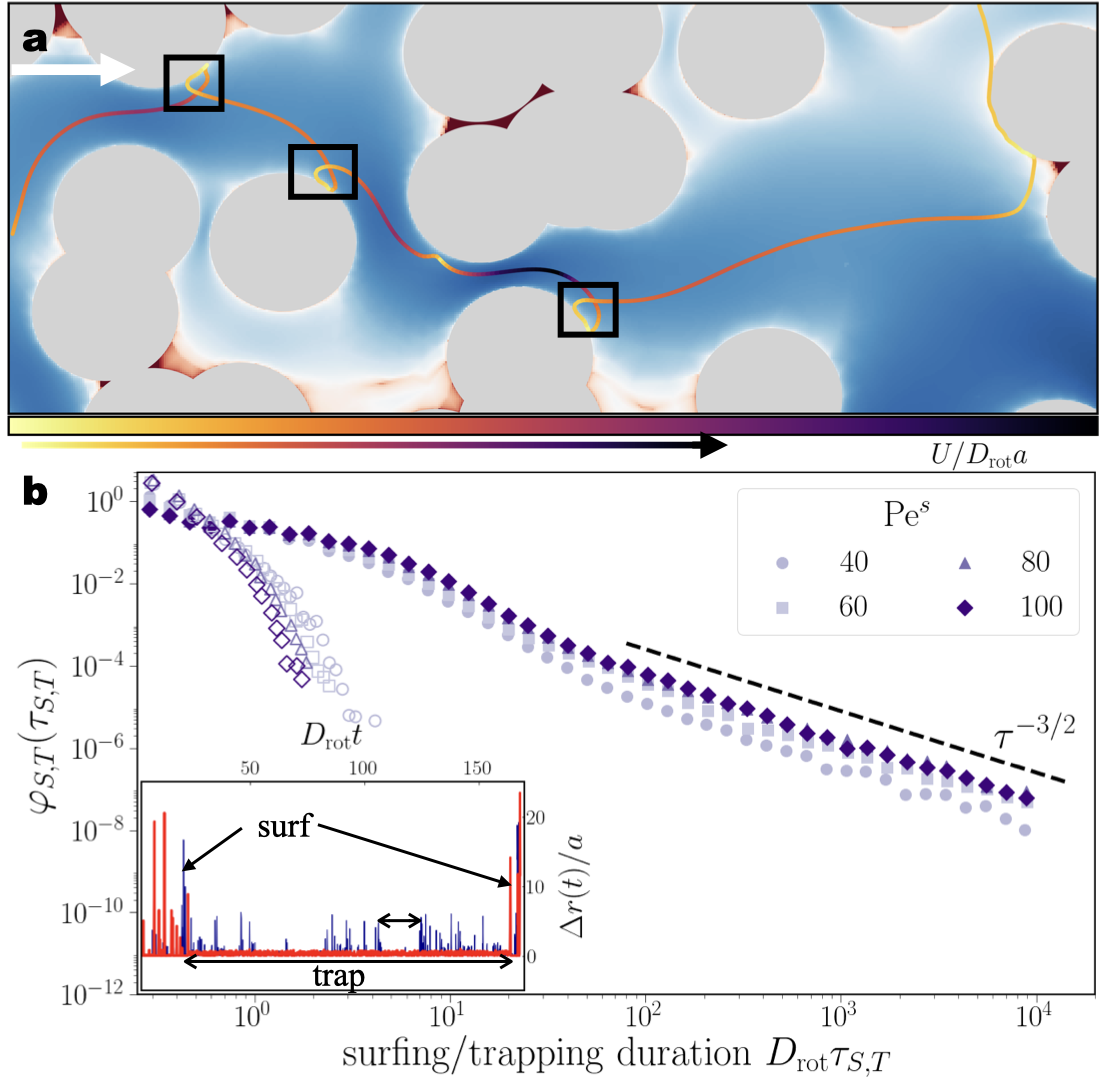}
\caption{ \label{fig:trap} {\bf Surf-and-trap motility. a.} Exemplary trajectory of a swimmer. Here, the activity is $\mathrm{Pe}^s=80$ and the agent's exit time is $\tau = 100/D_{\rm rot}$. The trajectory is colored according to the instantaneous particle velocity $U$. The color bar for the flow velocity is shown in Fig.~\ref{fig:method}. {\bf b.}~Trap- and surf-time distributions, $\varphi_T(\tau_T)$ and $\varphi_S(\tau_S)$, for different activities ${\rm Pe}^s$. ({\it Inset})~Exemplary displacements $\Delta r(t)$ as a function of time. Here, the packing fraction and flow P{\'e}clet number are $\phi = 0.51$ and $\mathrm{Pe}^f = 40$, respectively. We denote by $D_{\rm rot}$ the rotational diffusivity and $a$ is the obstacle radius.}
\end{figure}

\paragraph{Swimming slows down.--}We measure the distributions of the exit times $\varphi_{\rm exit}(\tau)$ for a range of activities ${\rm Pe}^s=0,\dots,100$ (Fig.~\ref{fig:exittime}{\bf a}). For passive Brownian particles (${\rm Pe}^s=0$), our results show a peaked distribution with a most probable exit time at $\tau^\star \approx 3/D_{\rm rot}$, which is similar to the distribution of the streamlines of the flow (black line in Fig.~\ref{fig:exittime}{\bf a}), indicating that the particles are mainly transported across the channel by the fluid flow. Upon increasing the activity ${\rm Pe}^s$, we find an overall shift of the peaks of the distributions to longer times, demonstrating a significant slowdown of transport with increasing activity. This result clearly shows that swimming prevents agents from rapidly crossing the porous channel.

Remarkably, this slowdown manifests in prominent  power-law tails of the exit-time distributions, $\varphi_{\rm exit}(\tau) \propto \tau^{-\alpha}$ with exponent $\alpha$, for  activities exceeding the flow P{\'e}clet number, $\mathrm{Pe}^s \gtrsim \mathrm{Pe}^f$. These can be collapsed onto a single master curve upon rescaling the exit times $\tau$ by the activity $\mathrm{Pe}^s$  and exhibit a scaling exponent of $\alpha \approx 3/2$ over three orders of magnitude in time (Fig.~\ref{fig:exittime}{\bf b}). This data collapse reveals that long exit times, $\tau \gtrsim \mathrm{Pe}^s / D_{\rm rot}$, are predominantly governed by the activity $\mathrm{Pe}^s$ of the active agents. Specifically, stronger self-propulsion or, equivalently, weaker rotational diffusion enhance directional persistence, causing agents to need more time to exit the porous media. In contrast, higher rotational diffusion induces more frequent reorientations, enabling swimmers to shorten their exit times. Overall, their exit times remain longer than those of their passive counterparts.

So far, we have examined this counter-intuitive behavior at a fixed packing fraction of the porous medium. A natural question then arises: Does this scaling persist at higher packing fractions, where crowding alters both flow fields and motility? Indeed, the long-time behavior exhibits a robust power-law decay (with $\alpha \approx 3/2$) and these long-time tails collapse onto a master curve across a wide range of packing fractions, $\phi = 0.34- 0.58$, in the regime $\mathrm{Pe}^s \gtrsim \mathrm{Pe}^f$ (Fig.~\ref{fig:exittime}{\bf c}). This data collapse demonstrates that, once the activity $\mathrm{Pe}^s$ and flow strength $\mathrm{Pe}^f$ are comparable and known, the exit-time statistics of active agents can be quantitatively predicted. 

The observed data collapse can be rationalized by a simple scaling argument. As the packing fraction increases, the mean flow velocity $\langle |\boldsymbol{u}| \rangle$ decreases (for a fixed pressure drop), leading to longer traversal times across the system. We can introduce two length scales: First, the distance between the inlet and outlet of the porous channel can be expressed as $L \sim \langle |\boldsymbol{u}| \rangle \tau$. Second,  the swimmer’s persistence length is $\ell_p = v / D_{\rm rot}$. The data collapse occurs when these characteristic length scales are comparable $L \sim \ell_p$, leading to $L/\ell_p\sim D_{\rm rot} \tau {\rm Pe}^f/{\rm Pe}^s$  as shown in Fig.~\ref{fig:exittime}{\bf c} (time axis).  

The appearance of the universal exponent $\alpha \approx 3/2$ evokes classic first-passage phenomena in comb-like geometries~\cite{bouchaud_anomalous_1990}, where a Brownian particle entering an infinitely long tooth experiences a broad, heavy-tailed distribution of waiting times before returning to and moving along the comb backbone. This arises because the probability that a one-dimensional Brownian walker returns to the origin of a comb tooth decays as $\propto \tau^{-3/2}$ and therefore rare, prolonged excursions dominate the long-time dynamics, giving rise to anomalous transport. We hypothesize that in our system an analogous mechanism is at play: active agents get trapped in the complex pore geometry (corresponding to the comb teeth) and can only leave these traps through rotational diffusion,  allowing them to reorient and swim away. This correspondence between the passive and active system is supported by the fact that the probability for an active particle to leave an open half-space through a boundary (similar to an infinitely large trap) via orientational diffusion behaves as $\propto \tau^{-3/2}$~\cite{Baouche:2025, Baouche:2025:arxiv}, akin to the passive case. 

\begin{figure*}
\centering
\includegraphics[width=1.0\linewidth]{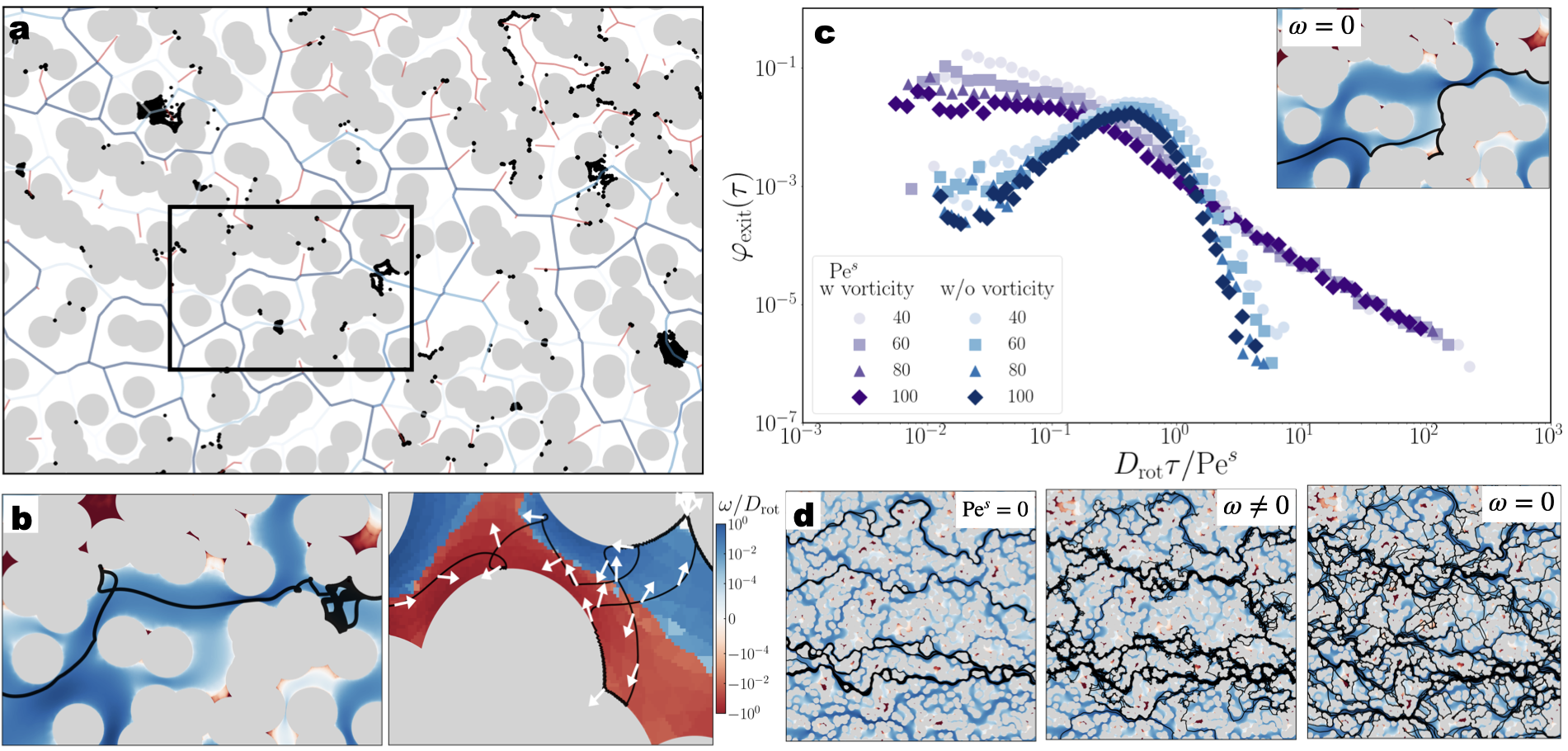}
\caption{ \label{fig:shear}{\bf Vorticity-induced trapping. a.}~Positions of active agents (black dots) during their trapping phases shown on top of the underlying flow network and porous geometry.  {\bf b.}~Trajectory and swimming orientation (arrows) of an active agent in the marked zone  with (left) fluid velocity and (right) fluid vorticity in the background. {\bf c.}~Rescaled exit-time distributions of swimmers with and without the effect of vorticity-induced reorientation.  ({\it Inset}) Trajectory of a swimmer without orientation-vorticity coupling. {\bf d.}~Trajectories of fastest~$10\%$ passive tracers (left), active agents with (middle) and without (right) orientation-vorticity coupling. Here, the flow P{\'e}clet number and packing fraction are $\mathrm{Pe}^f = 80$ and $\phi = 0.51$, respectively. Unless otherwise stated, the activity is $\mathrm{Pe}^s = 80$. We denote by $D_{\rm rot}$ the rotational diffusivity. The color bar for the flow velocity in {\bf a-d} is shown in Fig.~\ref{fig:method}. }
\end{figure*}

\paragraph{Surf-and-trap dynamics.--}To test our hypothesis and establish a firm understanding of the underlying physics, we closely examine the trajectories of agents with high activity and long exit times.  Figure~\ref{fig:trap}{\bf a} shows a representative trajectory, which is color-coded according to the agent's instantaneous velocity. Interestingly, the agent traverses several loop structures, which appear as a common feature for agents with high activities and are reminiscent of experimental trajectories of {\it Escherichia coli}~\cite{creppy_effect_2019}. Our results further indicate that agents move significantly slower and, consequently, spend more time in these looping parts of the trajectory than in the straight ones. 

We quantify these temporal differences in the swimmers' trajectories by analyzing their instantaneous displacements $\Delta r(t) = |\Delta \boldsymbol{r} (t)|$ at time $t$. Typically, these displacements show small fluctuations, interrupted by bursts of movement over brief intervals [Fig.~\ref{fig:trap}{\bf b}({\it inset})]. We differentiate these two phases of dynamics as the  `{\it trapping}' and `{\it surfing}' phases, respectively. We classify the trapping phase by the criterion $\Delta r(t)\leq\Delta r^{\rm free}/3$, where $\Delta r^{\rm free}$ represents the average displacement of the agent in free space. We note that our results are robust with respect to the cut-off. Using this approach, we measure the trapping times $\tau_T$, referred to the duration a swimmer spends in a trapping phase bounded by two surfing phases. Strikingly, we find that the trapping-time distribution  exhibits a power-law scaling $\varphi_T(\tau_T)\sim \tau_T^{-\alpha}$ for $\mathrm{Pe}^s \gtrsim \mathrm{Pe}^f$ with an exponent of $\alpha\approx3/2$, which aligns with the one observed for the exit-time distributions. 

Ultimately, upon encountering a fast flow path the agent surfs along the flow streamlines and traverses the channel. The associated surf times $\tau_S$ slightly decrease for higher activities, indicating that active motion speeds up transport across the channel during the surfing phase (Fig.~\ref{fig:trap}{\bf b}). An important feature of these dynamics is that the surfing phases are much shorter than the trapping phases~$\tau_S \ll \tau_T$ and therefore barely contribute to the long exit-times. Thus, overall these observations corroborate the picture of motion along a `comb potential'~\cite{bouchaud_anomalous_1990}: Active agents surf from one comb tooth to another, where the comb teeth correspond to traps with associated power-law trapping-time distributions [Fig.~\ref{fig:exittime}{\bf c}({\it inset})].

\paragraph{Vorticity-induced trapping.--}To reveal the physical mechanisms underlying the surf-and-trap dynamics, we study their relation to the spatial complexity of the crowded environment and the flow fields. The latter are highly-heterogeneous and can be coarsened in terms of a network representation, displaying two distinct features: the flow backbone, which carries most of the flow across the channel, and the dead ends, where the fluid cannot flow. 
By extracting the trapping locations, we find that agents become trapped at obstacle boundaries and dead ends (Fig.~\ref{fig:shear}{\bf a}). This is comparable to the accumulation of swimmers in a porous system without flow, as the dead ends have vanishingly low velocity. 

In addition and most strikingly, our results demonstrate that a significant amount of trapping occurs in the backbone of the pore space. Inspection of particle trajectories reveals that agents are trapped due to continuous reorientations with the flow vorticity, leading to motion against the flow direction. These features can result in cyclic orbits (Fig.~\ref{fig:shear}{\bf b}) until the agents align with the local flow direction. This vorticity-induced trapping strongly enforces the slow down of active agents in crowded media. 

To escape from traps the reorientation mechanism of active agents plays a crucial role ~\cite{Bhattacharjee:2019, Kurzthaler:2021}. In our model, the swimmer reorientation is determined by the interplay of fluid vorticity and rotational diffusion. To decouple these effects, we investigate the transport behavior of agents in the absence of orientation-vorticity coupling ($\omega = 0$). Figure~\ref{fig:shear}{\bf c}({\it inset}) depicts a representative trajectory, which displays a significantly different nature, where the swimmer either stays on the flow path or follows the obstacle boundaries. We do not observe any accumulation in the flow backbone, which is further reflected in the exit-time distributions (Fig.~\ref{fig:shear}{\bf c}). The latter decay faster at long times than those accounting for the orientation-vorticity coupling  and the long-time power-law tail does not manifest. This finding thus highlights the prominent role of fluid vorticity for the trapping of microswimmers and the resultant delayed transport. 

\paragraph{Vorticity promotes fast exits.--}Interestingly, we observe that the exit-time distributions also differ at early times (Fig.~\ref{fig:shear}{\bf c}). In particular, agents, which are not reoriented by the fluid vorticity, have their most probable exit time at $\tau^\star \sim {\rm Pe}^s/D_{\rm rot}$, which is significantly higher than that of their counterpart with $\tau^\star \sim 10^{-2}{\rm Pe}^s/D_{\rm rot}$. This phenomenon can be analyzed by comparing the trajectories of the `fastest' swimmers, which we define as the first~$10\%$ of agents exiting the channel. In the presence of orientation-vorticity coupling (Fig.~\ref{fig:shear}{\bf d}, middle panel), the swimmers are tracing specific flow paths, which are in part similar to those of their passive counterparts (Fig.~\ref{fig:shear}{\bf d}, left panel). However, in the absence of orientation-vorticity coupling (Fig.~\ref{fig:shear}{\bf c}, right panel),  the  agents spread through the system without any visible bias towards specific flow paths. Thus, fluid vorticity can aid swimmers in staying on the main flow path on a shorter time scale, thereby speeding up their traversal across the channel. 

\paragraph{Summary and conclusions.--} Our findings demonstrate the intricate physics arising from the coupling of the swimming direction of active agents and the highly-heterogeneous vorticity field of a disordered flow. Most importantly, we have identified that this orientation-vorticity coupling can capture active agents in the flow backbone, giving rise to long-lived trapping phases. We have further revealed a new `surf-and-trap' pattern, which underlies the power-law scaling of the exit-time distributions that remains robust across different packing fractions and motility parameters. We argue that the physical picture is reminiscent of diffusion along a `comb potential'~\cite{bouchaud_anomalous_1990}. In our system, vorticity-induced trapping plays the role of the comb teeth: swimmers are trapped within the porous medium for extended periods, escaping only via rotational diffusion before rejoining the main flow. This dynamic correspondence between the traverse of passive agents in a comb potential and that of active agents in vorticity-induced geometric traps explains the emergence of the same universal scaling, while highlighting a fundamentally new mechanism in which activity, pore geometry, and fluid flow collectively drive non-trivial transport in active systems.

The aspect of vorticity-induced trapping in the flow backbone is particularly important in broader contexts, such as microbiology and engineering, as it helps to understand bacterial invasion in various natural and synthetic environments, including porous soils or tissues~\cite{Jin:2024}. Notably, while we assume that swimmers do not impact the flow or interact with one another, our minimal model effectively replicates experimentally-observed dynamical patterns of microorganisms~\cite{creppy_effect_2019, Altshuler:2013} and highlights important biological phenomena such as clogging~\cite{Aufrecht:2019, Kurz:2022, Kurz:2023}. 

Future work should explore the generality of these transport mechanisms for different motility patterns~\cite{Licata:2016, Kurzthaler:2021}, the impact of swimmer shape~\cite{Barry:2015, Rusconi:2014, Vennamneni:2020}, and the interplay of heterogeneous flows on chemical fields and chemotaxis~\cite{DeAnna:2021, Grogno:2023}. Another relevant research direction constitutes the impact of a finite swimmer size on the dynamics at both the individual and collective level~\cite{Keogh:2024} in dense, disordered media; in particular, exploring how this behavior in turn impacts the velocity field~\cite{Residori:2023} and may induce clogging represents another important aspect. Finally, fluid shear appears to be omnipresent in microbial habitats, raising the question whether cells have evolved strategies to actively respond to shear and adapt their swimming behaviors.  

\section{Materials and Methods}
\paragraph{Model.--}  The governing equations for the instantaneous position $\boldsymbol{r}(t)$ and the orientation $\boldsymbol{e}(t)=[\cos\vartheta(t), \sin\vartheta(t)]$ of the agent read:
\begin{subequations}
\begin{align}
\frac{\mathrm{d} \boldsymbol{r}}{\mathrm{d}t} &= v \boldsymbol{e} +\boldsymbol{u}^{\rm HI} 
-\frac{1}{\zeta}\sum_{i=1}^N\nabla_i U+\sqrt{2D}\boldsymbol{\xi}, \label{eq:ABP1} \\ 
\frac{\mathrm{d} \vartheta }{\mathrm{d}t} &= \omega^{\rm HI} + \sqrt{2D_{\rm rot}}\chi,
\label{eq:ABP2}
\end{align}
\end{subequations}
where $\boldsymbol{u}^{\rm HI}$ and $\omega^{\rm HI}$ denote the contributions due to the flow field. We assume that the active agents are point particles, so that the hydrodynamic contributions can be expressed using Faxen's law~\cite{leal2007advanced}: 
\begin{subequations}
\begin{align}
\boldsymbol{u}^{\rm HI} &= \boldsymbol{u}(\boldsymbol{r}),\\
\omega^{\rm HI} &= \frac{1}{2} [\partial_x u_y(\boldsymbol{r})-\partial_y u_x(\boldsymbol{r})]=: \frac{\omega(\boldsymbol{r})}{2},
\end{align}
\end{subequations}
where $\omega(\boldsymbol{r})$ denotes the spatially-varying vorticity field.  For details of the computation of the latter we refer to the next section. Further, $\zeta$ is the frictional coefficient, and $\boldsymbol{\xi}(t)$ and $\chi(t)$ represent Gaussian white noise of zero mean and unit variance. We further use that the diffusion coefficients of spherical particles are related via the Stokes-Einstein relation $D/D_{\rm rot} = 4a_0^2/3$ with $a_0$ being the particle radius.

The porous geometry is modeled as a collection of $N$ randomly-placed overlapping discs of radius $a$  in a two dimensional domain, which is finite in the $x-$direction and periodic in the $y-$direction. The interaction of the agents with the obstacles is captured by a repulsive Weeks-Chandler-Andersen  potential:  
\begin{align}
U &= \begin{cases}  4 k_B T\epsilon\left[ \left(\frac{\sigma}{r_i}\right)^{12}- \left(\frac{\sigma}{r_{i}}\right)^6\right]+\epsilon& r_i < \sigma,\\
    0& r_i \geq \sigma,
    \end{cases} \label{eq:WCA}
\end{align} 
where $k_B$ denotes the Boltzmann constant and $T$ is the temperature. Further, $r_{i}$ is the distance between the microswimmer and the $i$th obstacle and we use $\sigma=a+a_0$ as effective radius composed of the swimmer and obstacle radii, mimicking hard-sphere-like interactions. The dimensionless strength of the interaction potential is $\epsilon$, which we choose as $\epsilon = 100$ throughout this work. The length of the porous domain is set to $L=150a$ and the ratio of particle-to-obstacle radius is $a_0/a =0.001$. For each packing fraction $\phi$, we obtain statistics for $20$ independent geometries and simulate $10^3$ trajectories for each of them over a maximum time of $10^4/D_{\rm rot}$. 

\paragraph{Velocity field.--}The spatially-dependent fluid velocity~$\boldsymbol{u}(\boldsymbol{r})$ and pressure field $p(\boldsymbol{r})$ in the porous channel are described by the Stokes equations:
\begin{align}
\mu \nabla^2\boldsymbol{u}=\nabla p
\quad {\rm and} \quad \nabla \cdot \boldsymbol{u} = 0, 
\end{align}
where  $\mu$ denotes fluid viscosity and a pressure drop $\Delta p$ is applied between the inlet and outlet of the channel. Imposing no-slip boundary conditions on the surface of the obstacles, the pressure and velocity fields are obtained using the finite-element method based on Taylor-Hood elements. Details can be found in in our previous work~\cite{Residori:2025}.

It is important to note that the solution of the Stokes equations is computed on an unstructured grid. To solve Eqs.~\eqref{eq:ABP1}-\eqref{eq:ABP2}, we resample the velocity field onto a structured rectangular grid, with square elements of dimension $\delta \approx 10^{-5}L$. The resampling procedure evaluates the velocity field at the center of the square and assigns that value to the whole element. The obstacle surface can cut the underlying elements. In that case, if the center of the element lies inside the obstacle surface, the resampled velocity would be zero on the whole square, with part of the element belonging to the pore space. This, in turn, would cause an artificial slowdown of agents with small activity traversing such elements. We avoid this artifact by replacing the velocity of the elements at the boundary with the averages of their neighboring elements. Even though this approach generates a small non-zero velocity at the boundary, contradicting the no-slip boundary condition, the above-mentioned repulsive potential prevents the agents from colliding with the obstacles. We perform the same smoothing procedure when the velocity of an element is zero and changes to a nonzero value at any of its neighboring elements beyond a cut-off. This ensures that no abrupt velocity change occurs around the obstacle boundary or in extremely low flow zones. The same modifications are applied for the vorticity field. We validate this approach by comparing the exit-time distributions of passive tracers (${\rm Pe}^s=0$) using both the unstructured grid (black line) and the rectangular grid in Fig.~\ref {fig:exittime}{\bf a}, showing good agreement.

To coarsen the complex, heterogeneous flow field, we derive a network representation of the flow through the dense porous medium following Ref.~\cite{elam_1984}. We construct the Voronoi tessellation, taking the centers of the obstacles as points. We remove the Voronoi edges intersecting any obstacle and, for the remaining edges, compute the distance $d_{ij}$ from the edge to the closest obstacle. We rely on the lubrication theory~\cite{leal2007advanced} to assign a permeability~$k_{ij}$ to each edge. In particular, we approximate the pore space between two obstacles corresponding to a Voronoi edge with a pipe of height $2d_{ij}$ and length $(2d_{ij} a)^{1/2}$. The permeability is proportional to $d_{ij}^{5/2}$, which corresponds to the weight assigned to the Voronoi edge. Finally, we add two nodes to the network -- a source and a sink -- and compute the pressure $p_i$ on each node and the flow on each edge $k_{ij} (p_i - p_j)$ by solving $q_i = \sum_{j} (p_j - p_i) k_{ij}$ for all $i$, where $q_i$ corresponds to the flow rate.

\section{Data availability}
The data are available from the corresponding author upon reasonable request.
\section{Code availability}
The computer code used for simulations is available from the corresponding author upon reasonable request.
\section{Acknowledgments}
P.D. acknowledges funding from the Alexander von Humboldt-Stiftung. A.V. acknowledges funding from the German Research Foundation (DFG) – Project number: 556185784. We are further grateful for discussions with Akhil Varma, Yanis Baouche, and Sagnik Garai. 

\bibliography{literature}

\begin{thebibliography}{62}%
\makeatletter
\providecommand \@ifxundefined [1]{%
 \@ifx{#1\undefined}
}%
\providecommand \@ifnum [1]{%
 \ifnum #1\expandafter \@firstoftwo
 \else \expandafter \@secondoftwo
 \fi
}%
\providecommand \@ifx [1]{%
 \ifx #1\expandafter \@firstoftwo
 \else \expandafter \@secondoftwo
 \fi
}%
\providecommand \natexlab [1]{#1}%
\providecommand \enquote  [1]{``#1''}%
\providecommand \bibnamefont  [1]{#1}%
\providecommand \bibfnamefont [1]{#1}%
\providecommand \citenamefont [1]{#1}%
\providecommand \href@noop [0]{\@secondoftwo}%
\providecommand \href [0]{\begingroup \@sanitize@url \@href}%
\providecommand \@href[1]{\@@startlink{#1}\@@href}%
\providecommand \@@href[1]{\endgroup#1\@@endlink}%
\providecommand \@sanitize@url [0]{\catcode `\\12\catcode `\$12\catcode
  `\&12\catcode `\#12\catcode `\^12\catcode `\_12\catcode `\%12\relax}%
\providecommand \@@startlink[1]{}%
\providecommand \@@endlink[0]{}%
\providecommand \url  [0]{\begingroup\@sanitize@url \@url }%
\providecommand \@url [1]{\endgroup\@href {#1}{\urlprefix }}%
\providecommand \urlprefix  [0]{URL }%
\providecommand \Eprint [0]{\href }%
\providecommand \doibase [0]{https://doi.org/}%
\providecommand \selectlanguage [0]{\@gobble}%
\providecommand \bibinfo  [0]{\@secondoftwo}%
\providecommand \bibfield  [0]{\@secondoftwo}%
\providecommand \translation [1]{[#1]}%
\providecommand \BibitemOpen [0]{}%
\providecommand \bibitemStop [0]{}%
\providecommand \bibitemNoStop [0]{.\EOS\space}%
\providecommand \EOS [0]{\spacefactor3000\relax}%
\providecommand \BibitemShut  [1]{\csname bibitem#1\endcsname}%
\let\auto@bib@innerbib\@empty
\bibitem [{\citenamefont {Tecon}\ and\ \citenamefont {Or}(2017)}]{Tecon:2017}%
  \BibitemOpen
  \bibfield  {author} {\bibinfo {author} {\bibfnamefont {R.}~\bibnamefont
  {Tecon}}\ and\ \bibinfo {author} {\bibfnamefont {D.}~\bibnamefont {Or}},\
  }\bibfield  {title} {\bibinfo {title} {Biophysical processes supporting the
  diversity of microbial life in soil},\ }\href
  {https://doi.org/10.1093/femsre/fux039} {\bibfield  {journal} {\bibinfo
  {journal} {FEMS Microbiology Reviews}\ }\textbf {\bibinfo {volume} {41}},\
  \bibinfo {pages} {599–623} (\bibinfo {year} {2017})}\BibitemShut {NoStop}%
\bibitem [{\citenamefont {Voigtländer}\ \emph {et~al.}(2024)\citenamefont
  {Voigtländer}, \citenamefont {Houssais}, \citenamefont {Bacik},
  \citenamefont {Bourg}, \citenamefont {Burton}, \citenamefont {Daniels},
  \citenamefont {Datta}, \citenamefont {Gado}, \citenamefont {Deshpande},
  \citenamefont {Devauchelle} \emph {et~al.}}]{Voigtlaender:2023}%
  \BibitemOpen
  \bibfield  {author} {\bibinfo {author} {\bibfnamefont {A.}~\bibnamefont
  {Voigtländer}}, \bibinfo {author} {\bibfnamefont {M.}~\bibnamefont
  {Houssais}}, \bibinfo {author} {\bibfnamefont {K.~A.}\ \bibnamefont {Bacik}},
  \bibinfo {author} {\bibfnamefont {I.~C.}\ \bibnamefont {Bourg}}, \bibinfo
  {author} {\bibfnamefont {J.~C.}\ \bibnamefont {Burton}}, \bibinfo {author}
  {\bibfnamefont {K.~E.}\ \bibnamefont {Daniels}}, \bibinfo {author}
  {\bibfnamefont {S.~S.}\ \bibnamefont {Datta}}, \bibinfo {author}
  {\bibfnamefont {E.~D.}\ \bibnamefont {Gado}}, \bibinfo {author}
  {\bibfnamefont {N.~S.}\ \bibnamefont {Deshpande}}, \bibinfo {author}
  {\bibfnamefont {O.}~\bibnamefont {Devauchelle}}, \emph {et~al.},\ }\bibfield
  {title} {\bibinfo {title} {Soft matter physics of the ground beneath our
  feet},\ }\href {https://doi.org/10.1039/D4SM00391H} {\bibfield  {journal}
  {\bibinfo  {journal} {Soft Matter}\ }\textbf {\bibinfo {volume} {20}},\
  \bibinfo {pages} {5859–5888} (\bibinfo {year} {2024})}\BibitemShut
  {NoStop}%
\bibitem [{\citenamefont {Ó~Cróinín}\ and\ \citenamefont
  {Backert}(2012)}]{Croinin:2012}%
  \BibitemOpen
  \bibfield  {author} {\bibinfo {author} {\bibfnamefont {T.}~\bibnamefont
  {Ó~Cróinín}}\ and\ \bibinfo {author} {\bibfnamefont {S.}~\bibnamefont
  {Backert}},\ }\bibfield  {title} {\bibinfo {title} {Host epithelial cell
  invasion by campylobacter jejuni: Trigger or zipper mechanism?},\ }\href
  {https://doi.org/10.3389/fcimb.2012.00025} {\bibfield  {journal} {\bibinfo
  {journal} {Frontiers in Cellular and Infection Microbiology}\ }\textbf
  {\bibinfo {volume} {2}},\ \bibinfo {pages} {1} (\bibinfo {year}
  {2012})}\BibitemShut {NoStop}%
\bibitem [{\citenamefont {Kirsch}\ \emph {et~al.}(2019)\citenamefont {Kirsch},
  \citenamefont {Basche}, \citenamefont {Neunzehn}, \citenamefont {Dede},
  \citenamefont {Dannemann}, \citenamefont {Hannig},\ and\ \citenamefont
  {Weber}}]{Kirsch:2019}%
  \BibitemOpen
  \bibfield  {author} {\bibinfo {author} {\bibfnamefont {J.}~\bibnamefont
  {Kirsch}}, \bibinfo {author} {\bibfnamefont {S.}~\bibnamefont {Basche}},
  \bibinfo {author} {\bibfnamefont {J.}~\bibnamefont {Neunzehn}}, \bibinfo
  {author} {\bibfnamefont {M.}~\bibnamefont {Dede}}, \bibinfo {author}
  {\bibfnamefont {M.}~\bibnamefont {Dannemann}}, \bibinfo {author}
  {\bibfnamefont {C.}~\bibnamefont {Hannig}},\ and\ \bibinfo {author}
  {\bibfnamefont {M.-T.}\ \bibnamefont {Weber}},\ }\bibfield  {title} {\bibinfo
  {title} {Is it really penetration? {P}art 2. {L}ocomotion of {E}nterococcus
  faecalis cells within dentinal tubules of bovine teeth},\ }\href
  {https://doi.org/10.1007/s00784-019-02865-5} {\bibfield  {journal} {\bibinfo
  {journal} {Clinical Oral Investigations}\ }\textbf {\bibinfo {volume} {23}},\
  \bibinfo {pages} {4325–4334} (\bibinfo {year} {2019})}\BibitemShut
  {NoStop}%
\bibitem [{\citenamefont {Otte}\ \emph {et~al.}(2021)\citenamefont {Otte},
  \citenamefont {Ipiña}, \citenamefont {Pontier-Bres}, \citenamefont
  {Czerucka},\ and\ \citenamefont {Peruani}}]{Otte:2021}%
  \BibitemOpen
  \bibfield  {author} {\bibinfo {author} {\bibfnamefont {S.}~\bibnamefont
  {Otte}}, \bibinfo {author} {\bibfnamefont {E.~P.}\ \bibnamefont {Ipiña}},
  \bibinfo {author} {\bibfnamefont {R.}~\bibnamefont {Pontier-Bres}}, \bibinfo
  {author} {\bibfnamefont {D.}~\bibnamefont {Czerucka}},\ and\ \bibinfo
  {author} {\bibfnamefont {F.}~\bibnamefont {Peruani}},\ }\bibfield  {title}
  {\bibinfo {title} {Statistics of pathogenic bacteria in the search of host
  cells},\ }\href {https://doi.org/10.1038/s41467-021-22156-6} {\bibfield
  {journal} {\bibinfo  {journal} {Nature Communications}\ }\textbf {\bibinfo
  {volume} {12}},\ \bibinfo {pages} {1990} (\bibinfo {year}
  {2021})}\BibitemShut {NoStop}%
\bibitem [{\citenamefont {Suarez}\ and\ \citenamefont
  {Pacey}(2006)}]{Suarez:2006}%
  \BibitemOpen
  \bibfield  {author} {\bibinfo {author} {\bibfnamefont {S.}~\bibnamefont
  {Suarez}}\ and\ \bibinfo {author} {\bibfnamefont {A.~A.}\ \bibnamefont
  {Pacey}},\ }\bibfield  {title} {\bibinfo {title} {Sperm transport in the
  female reproductive tract},\ }\href {https://doi.org/10.1093/humupd/dmi047}
  {\bibfield  {journal} {\bibinfo  {journal} {Human Reproduction Update}\
  }\textbf {\bibinfo {volume} {12}},\ \bibinfo {pages} {23–37} (\bibinfo
  {year} {2006})}\BibitemShut {NoStop}%
\bibitem [{\citenamefont {Hartmann}\ \emph {et~al.}(2019)\citenamefont
  {Hartmann}, \citenamefont {Singh}, \citenamefont {Pearce}, \citenamefont
  {Mok}, \citenamefont {Song}, \citenamefont {Díaz-Pascual}, \citenamefont
  {Dunkel},\ and\ \citenamefont {Drescher}}]{Hartmann:2019}%
  \BibitemOpen
  \bibfield  {author} {\bibinfo {author} {\bibfnamefont {R.}~\bibnamefont
  {Hartmann}}, \bibinfo {author} {\bibfnamefont {P.~K.}\ \bibnamefont {Singh}},
  \bibinfo {author} {\bibfnamefont {P.}~\bibnamefont {Pearce}}, \bibinfo
  {author} {\bibfnamefont {R.}~\bibnamefont {Mok}}, \bibinfo {author}
  {\bibfnamefont {B.}~\bibnamefont {Song}}, \bibinfo {author} {\bibfnamefont
  {F.}~\bibnamefont {Díaz-Pascual}}, \bibinfo {author} {\bibfnamefont
  {J.}~\bibnamefont {Dunkel}},\ and\ \bibinfo {author} {\bibfnamefont
  {K.}~\bibnamefont {Drescher}},\ }\bibfield  {title} {\bibinfo {title}
  {Emergence of three-dimensional order and structure in growing biofilms},\
  }\href {https://doi.org/10.1038/s41567-018-0356-9} {\bibfield  {journal}
  {\bibinfo  {journal} {Nature Physics}\ }\textbf {\bibinfo {volume} {15}},\
  \bibinfo {pages} {251–256} (\bibinfo {year} {2019})}\BibitemShut {NoStop}%
\bibitem [{\citenamefont {Hallatschek}\ \emph {et~al.}(2023)\citenamefont
  {Hallatschek}, \citenamefont {Datta}, \citenamefont {Drescher}, \citenamefont
  {Dunkel}, \citenamefont {Elgeti}, \citenamefont {Waclaw},\ and\ \citenamefont
  {Wingreen}}]{Hallatschek:2023}%
  \BibitemOpen
  \bibfield  {author} {\bibinfo {author} {\bibfnamefont {O.}~\bibnamefont
  {Hallatschek}}, \bibinfo {author} {\bibfnamefont {S.~S.}\ \bibnamefont
  {Datta}}, \bibinfo {author} {\bibfnamefont {K.}~\bibnamefont {Drescher}},
  \bibinfo {author} {\bibfnamefont {J.}~\bibnamefont {Dunkel}}, \bibinfo
  {author} {\bibfnamefont {J.}~\bibnamefont {Elgeti}}, \bibinfo {author}
  {\bibfnamefont {B.}~\bibnamefont {Waclaw}},\ and\ \bibinfo {author}
  {\bibfnamefont {N.~S.}\ \bibnamefont {Wingreen}},\ }\bibfield  {title}
  {\bibinfo {title} {Proliferating active matter},\ }\href
  {https://doi.org/10.1038/s42254-023-00593-0} {\bibfield  {journal} {\bibinfo
  {journal} {Nature Reviews Physics}\ }\textbf {\bibinfo {volume} {5}},\
  \bibinfo {pages} {407–419} (\bibinfo {year} {2023})}\BibitemShut {NoStop}%
\bibitem [{\citenamefont {Seymour}\ \emph {et~al.}(2017)\citenamefont
  {Seymour}, \citenamefont {Amin}, \citenamefont {Raina},\ and\ \citenamefont
  {Stocker}}]{Seymour:2017}%
  \BibitemOpen
  \bibfield  {author} {\bibinfo {author} {\bibfnamefont {J.~R.}\ \bibnamefont
  {Seymour}}, \bibinfo {author} {\bibfnamefont {S.~A.}\ \bibnamefont {Amin}},
  \bibinfo {author} {\bibfnamefont {J.-B.}\ \bibnamefont {Raina}},\ and\
  \bibinfo {author} {\bibfnamefont {R.}~\bibnamefont {Stocker}},\ }\bibfield
  {title} {\bibinfo {title} {Zooming in on the phycosphere: the ecological
  interface for phytoplankton–bacteria relationships},\ }\href
  {https://doi.org/10.1038/nmicrobiol.2017.65} {\bibfield  {journal} {\bibinfo
  {journal} {Nature Microbiology}\ }\textbf {\bibinfo {volume} {2}},\ \bibinfo
  {pages} {17065} (\bibinfo {year} {2017})}\BibitemShut {NoStop}%
\bibitem [{\citenamefont {Nguyen}\ \emph {et~al.}(2021)\citenamefont {Nguyen},
  \citenamefont {Lara-Gutiérrez},\ and\ \citenamefont
  {Stocker}}]{Nguyen:2021}%
  \BibitemOpen
  \bibfield  {author} {\bibinfo {author} {\bibfnamefont {J.}~\bibnamefont
  {Nguyen}}, \bibinfo {author} {\bibfnamefont {J.}~\bibnamefont
  {Lara-Gutiérrez}},\ and\ \bibinfo {author} {\bibfnamefont {R.}~\bibnamefont
  {Stocker}},\ }\bibfield  {title} {\bibinfo {title} {Environmental
  fluctuations and their effects on microbial communities, populations and
  individuals},\ }\href {https://doi.org/10.1093/femsre/fuaa068} {\bibfield
  {journal} {\bibinfo  {journal} {FEMS Microbiology Reviews}\ }\textbf
  {\bibinfo {volume} {45}},\ \bibinfo {pages} {fuaa068} (\bibinfo {year}
  {2021})}\BibitemShut {NoStop}%
\bibitem [{\citenamefont {Erkoc}\ \emph {et~al.}(2019)\citenamefont {Erkoc},
  \citenamefont {Yasa}, \citenamefont {Ceylan}, \citenamefont {Yasa},
  \citenamefont {Alapan},\ and\ \citenamefont {Sitti}}]{Erkoc:2019}%
  \BibitemOpen
  \bibfield  {author} {\bibinfo {author} {\bibfnamefont {P.}~\bibnamefont
  {Erkoc}}, \bibinfo {author} {\bibfnamefont {I.~C.}\ \bibnamefont {Yasa}},
  \bibinfo {author} {\bibfnamefont {H.}~\bibnamefont {Ceylan}}, \bibinfo
  {author} {\bibfnamefont {O.}~\bibnamefont {Yasa}}, \bibinfo {author}
  {\bibfnamefont {Y.}~\bibnamefont {Alapan}},\ and\ \bibinfo {author}
  {\bibfnamefont {M.}~\bibnamefont {Sitti}},\ }\bibfield  {title} {\bibinfo
  {title} {Mobile microrobots for active therapeutic delivery},\ }\href
  {https://doi.org/10.1002/adtp.201800064} {\bibfield  {journal} {\bibinfo
  {journal} {Advanced Therapeutics}\ }\textbf {\bibinfo {volume} {2}},\
  \bibinfo {pages} {1800064} (\bibinfo {year} {2019})}\BibitemShut {NoStop}%
\bibitem [{\citenamefont {Sedighi}\ \emph {et~al.}(2019)\citenamefont
  {Sedighi}, \citenamefont {Zahedi~Bialvaei}, \citenamefont {Hamblin},
  \citenamefont {Ohadi}, \citenamefont {Asadi}, \citenamefont {Halajzadeh},
  \citenamefont {Lohrasbi}, \citenamefont {Mohammadzadeh}, \citenamefont
  {Amiriani}, \citenamefont {Krutova} \emph {et~al.}}]{Sedighi:2019}%
  \BibitemOpen
  \bibfield  {author} {\bibinfo {author} {\bibfnamefont {M.}~\bibnamefont
  {Sedighi}}, \bibinfo {author} {\bibfnamefont {A.}~\bibnamefont
  {Zahedi~Bialvaei}}, \bibinfo {author} {\bibfnamefont {M.~R.}\ \bibnamefont
  {Hamblin}}, \bibinfo {author} {\bibfnamefont {E.}~\bibnamefont {Ohadi}},
  \bibinfo {author} {\bibfnamefont {A.}~\bibnamefont {Asadi}}, \bibinfo
  {author} {\bibfnamefont {M.}~\bibnamefont {Halajzadeh}}, \bibinfo {author}
  {\bibfnamefont {V.}~\bibnamefont {Lohrasbi}}, \bibinfo {author}
  {\bibfnamefont {N.}~\bibnamefont {Mohammadzadeh}}, \bibinfo {author}
  {\bibfnamefont {T.}~\bibnamefont {Amiriani}}, \bibinfo {author}
  {\bibfnamefont {M.}~\bibnamefont {Krutova}}, \emph {et~al.},\ }\bibfield
  {title} {\bibinfo {title} {Therapeutic bacteria to combat cancer; current
  advances, challenges, and opportunities},\ }\href
  {https://doi.org/10.1002/cam4.2148} {\bibfield  {journal} {\bibinfo
  {journal} {Cancer Medicine}\ }\textbf {\bibinfo {volume} {8}},\ \bibinfo
  {pages} {3167} (\bibinfo {year} {2019})}\BibitemShut {NoStop}%
\bibitem [{\citenamefont {Gao}\ and\ \citenamefont {Wang}(2014)}]{Gao:2014}%
  \BibitemOpen
  \bibfield  {author} {\bibinfo {author} {\bibfnamefont {W.}~\bibnamefont
  {Gao}}\ and\ \bibinfo {author} {\bibfnamefont {J.}~\bibnamefont {Wang}},\
  }\bibfield  {title} {\bibinfo {title} {The environmental impact of
  micro/nanomachines: A review},\ }\href {https://doi.org/10.1021/nn500077a}
  {\bibfield  {journal} {\bibinfo  {journal} {ACS Nano}\ }\textbf {\bibinfo
  {volume} {8}},\ \bibinfo {pages} {3170} (\bibinfo {year} {2014})}\BibitemShut
  {NoStop}%
\bibitem [{\citenamefont {Adadevoh}\ \emph {et~al.}(2016)\citenamefont
  {Adadevoh}, \citenamefont {Triolo}, \citenamefont {Ramsburg},\ and\
  \citenamefont {Ford}}]{Adadevoh:2016}%
  \BibitemOpen
  \bibfield  {author} {\bibinfo {author} {\bibfnamefont {J.~S.~T.}\
  \bibnamefont {Adadevoh}}, \bibinfo {author} {\bibfnamefont {S.}~\bibnamefont
  {Triolo}}, \bibinfo {author} {\bibfnamefont {C.~A.}\ \bibnamefont
  {Ramsburg}},\ and\ \bibinfo {author} {\bibfnamefont {R.~M.}\ \bibnamefont
  {Ford}},\ }\bibfield  {title} {\bibinfo {title} {Chemotaxis increases the
  residence time of bacteria in granular media containing distributed
  contaminant sources},\ }\href {https://doi.org/10.1021/acs.est.5b03956}
  {\bibfield  {journal} {\bibinfo  {journal} {Environmental Science and
  Technology}\ }\textbf {\bibinfo {volume} {50}},\ \bibinfo {pages} {181}
  (\bibinfo {year} {2016})}\BibitemShut {NoStop}%
\bibitem [{\citenamefont {Li}\ \emph {et~al.}(2017)\citenamefont {Li},
  \citenamefont {Esteban-Fern{\'a}ndez~de {\'A}vila}, \citenamefont {Gao},
  \citenamefont {Zhang},\ and\ \citenamefont {Wang}}]{Li:2017}%
  \BibitemOpen
  \bibfield  {author} {\bibinfo {author} {\bibfnamefont {J.}~\bibnamefont
  {Li}}, \bibinfo {author} {\bibfnamefont {B.}~\bibnamefont
  {Esteban-Fern{\'a}ndez~de {\'A}vila}}, \bibinfo {author} {\bibfnamefont
  {W.}~\bibnamefont {Gao}}, \bibinfo {author} {\bibfnamefont {L.}~\bibnamefont
  {Zhang}},\ and\ \bibinfo {author} {\bibfnamefont {J.}~\bibnamefont {Wang}},\
  }\bibfield  {title} {\bibinfo {title} {Micro/nanorobots for biomedicine:
  Delivery, surgery, sensing, and detoxification},\ }\href
  {https://doi.org/10.1126/scirobotics.aam6431} {\bibfield  {journal} {\bibinfo
   {journal} {Science Robotics}\ }\textbf {\bibinfo {volume} {2}},\ \bibinfo
  {pages} {eaam6431} (\bibinfo {year} {2017})}\BibitemShut {NoStop}%
\bibitem [{\citenamefont {de~Carvalho}(2018)}]{Carvalho:2018}%
  \BibitemOpen
  \bibfield  {author} {\bibinfo {author} {\bibfnamefont {C.~C. C.~R.}\
  \bibnamefont {de~Carvalho}},\ }\bibfield  {title} {\bibinfo {title} {Marine
  biofilms: A successful microbial strategy with economic implications},\
  }\href {https://doi.org/10.3389/fmars.2018.00126} {\bibfield  {journal}
  {\bibinfo  {journal} {Frontiers in Marine Sciences}\ }\textbf {\bibinfo
  {volume} {5}},\ \bibinfo {pages} {126} (\bibinfo {year} {2018})}\BibitemShut
  {NoStop}%
\bibitem [{\citenamefont {Kurzthaler}\ \emph {et~al.}(2023)\citenamefont
  {Kurzthaler}, \citenamefont {Gentile},\ and\ \citenamefont
  {Stone}}]{Kurzthaler:2023}%
  \BibitemOpen
  \bibfield  {author} {\bibinfo {author} {\bibfnamefont {C.}~\bibnamefont
  {Kurzthaler}}, \bibinfo {author} {\bibfnamefont {L.}~\bibnamefont
  {Gentile}},\ and\ \bibinfo {author} {\bibfnamefont {H.~A.}\ \bibnamefont
  {Stone}},\ }\href {https://doi.org/10.1039/9781839169465} {\emph {\bibinfo
  {title} {{Out-of-equilibrium Soft Matter}}}}\ (\bibinfo  {publisher} {The
  Royal Society of Chemistry},\ \bibinfo {year} {2023})\BibitemShut {NoStop}%
\bibitem [{\citenamefont {Kumar}\ \emph {et~al.}(2022)\citenamefont {Kumar},
  \citenamefont {Guasto},\ and\ \citenamefont {Ardekani}}]{Kumar:2022}%
  \BibitemOpen
  \bibfield  {author} {\bibinfo {author} {\bibfnamefont {M.}~\bibnamefont
  {Kumar}}, \bibinfo {author} {\bibfnamefont {J.~S.}\ \bibnamefont {Guasto}},\
  and\ \bibinfo {author} {\bibfnamefont {A.~M.}\ \bibnamefont {Ardekani}},\
  }\bibfield  {title} {\bibinfo {title} {Transport of complex and active fluids
  in porous media},\ }\href {https://doi.org/10.1122/8.0000389} {\bibfield
  {journal} {\bibinfo  {journal} {Journal of Rheology}\ }\textbf {\bibinfo
  {volume} {66}},\ \bibinfo {pages} {375–397} (\bibinfo {year}
  {2022})}\BibitemShut {NoStop}%
\bibitem [{\citenamefont {Spagnolie}\ and\ \citenamefont
  {Underhill}(2023)}]{Spagnolie:2023}%
  \BibitemOpen
  \bibfield  {author} {\bibinfo {author} {\bibfnamefont {S.~E.}\ \bibnamefont
  {Spagnolie}}\ and\ \bibinfo {author} {\bibfnamefont {P.~T.}\ \bibnamefont
  {Underhill}},\ }\bibfield  {title} {\bibinfo {title} {Swimming in complex
  fluids},\ }\href {https://doi.org/10.1146/annurev-conmatphys-040821-112149}
  {\bibfield  {journal} {\bibinfo  {journal} {Annual Review of Condensed Matter
  Physics}\ }\textbf {\bibinfo {volume} {14}},\ \bibinfo {pages} {381–415}
  (\bibinfo {year} {2023})}\BibitemShut {NoStop}%
\bibitem [{\citenamefont {Jin}\ and\ \citenamefont
  {Sengupta}(2024)}]{Jin:2024}%
  \BibitemOpen
  \bibfield  {author} {\bibinfo {author} {\bibfnamefont {C.}~\bibnamefont
  {Jin}}\ and\ \bibinfo {author} {\bibfnamefont {A.}~\bibnamefont {Sengupta}},\
  }\bibfield  {title} {\bibinfo {title} {Microbes in porous environments: from
  active interactions to emergent feedback},\ }\href
  {https://doi.org/10.1007/s12551-024-01185-7} {\bibfield  {journal} {\bibinfo
  {journal} {Biophysical Reviews}\ }\textbf {\bibinfo {volume} {16}},\ \bibinfo
  {pages} {173–188} (\bibinfo {year} {2024})}\BibitemShut {NoStop}%
\bibitem [{\citenamefont {Guasto}\ \emph {et~al.}(2012)\citenamefont {Guasto},
  \citenamefont {Rusconi},\ and\ \citenamefont {Stocker}}]{Guasto:2012}%
  \BibitemOpen
  \bibfield  {author} {\bibinfo {author} {\bibfnamefont {J.~S.}\ \bibnamefont
  {Guasto}}, \bibinfo {author} {\bibfnamefont {R.}~\bibnamefont {Rusconi}},\
  and\ \bibinfo {author} {\bibfnamefont {R.}~\bibnamefont {Stocker}},\
  }\bibfield  {title} {\bibinfo {title} {Fluid mechanics of planktonic
  microorganisms},\ }\href
  {https://doi.org/10.1146/annurev-fluid-120710-101156} {\bibfield  {journal}
  {\bibinfo  {journal} {Annual Review of Fluid Mechanics}\ }\textbf {\bibinfo
  {volume} {44}},\ \bibinfo {pages} {373–400} (\bibinfo {year}
  {2012})}\BibitemShut {NoStop}%
\bibitem [{\citenamefont {Wheeler}\ \emph {et~al.}(2019)\citenamefont
  {Wheeler}, \citenamefont {Secchi}, \citenamefont {Rusconi},\ and\
  \citenamefont {Stocker}}]{Wheeler:2019}%
  \BibitemOpen
  \bibfield  {author} {\bibinfo {author} {\bibfnamefont {J.~D.}\ \bibnamefont
  {Wheeler}}, \bibinfo {author} {\bibfnamefont {E.}~\bibnamefont {Secchi}},
  \bibinfo {author} {\bibfnamefont {R.}~\bibnamefont {Rusconi}},\ and\ \bibinfo
  {author} {\bibfnamefont {R.}~\bibnamefont {Stocker}},\ }\bibfield  {title}
  {\bibinfo {title} {Not just going with the flow: The effects of fluid flow on
  bacteria and plankton},\ }\href
  {https://doi.org/10.1146/annurev-cellbio-100818-125119} {\bibfield  {journal}
  {\bibinfo  {journal} {Annual Review of Cell and Developmental Biology}\
  }\textbf {\bibinfo {volume} {35}},\ \bibinfo {pages} {213–237} (\bibinfo
  {year} {2019})}\BibitemShut {NoStop}%
\bibitem [{\citenamefont {Mathijssen}\ \emph {et~al.}(2019)\citenamefont
  {Mathijssen}, \citenamefont {Figueroa-Morales}, \citenamefont {Junot},
  \citenamefont {Cl{\'e}ment}, \citenamefont {Lindner},\ and\ \citenamefont
  {Z{\"o}ttl}}]{Mathijssen:2019}%
  \BibitemOpen
  \bibfield  {author} {\bibinfo {author} {\bibfnamefont {A.~J. T.~M.}\
  \bibnamefont {Mathijssen}}, \bibinfo {author} {\bibfnamefont
  {N.}~\bibnamefont {Figueroa-Morales}}, \bibinfo {author} {\bibfnamefont
  {G.}~\bibnamefont {Junot}}, \bibinfo {author} {\bibfnamefont
  {{\'E}.}~\bibnamefont {Cl{\'e}ment}}, \bibinfo {author} {\bibfnamefont
  {A.}~\bibnamefont {Lindner}},\ and\ \bibinfo {author} {\bibfnamefont
  {A.}~\bibnamefont {Z{\"o}ttl}},\ }\bibfield  {title} {\bibinfo {title}
  {Oscillatory surface rheotaxis of swimming {E}. coli bacteria},\ }\href
  {https://doi.org/10.1038/s41467-019-11360-0} {\bibfield  {journal} {\bibinfo
  {journal} {Nature Communications}\ }\textbf {\bibinfo {volume} {10}},\
  \bibinfo {pages} {3434} (\bibinfo {year} {2019})}\BibitemShut {NoStop}%
\bibitem [{\citenamefont {Peng}\ and\ \citenamefont {Brady}(2020)}]{Peng:2020}%
  \BibitemOpen
  \bibfield  {author} {\bibinfo {author} {\bibfnamefont {Z.}~\bibnamefont
  {Peng}}\ and\ \bibinfo {author} {\bibfnamefont {J.~F.}\ \bibnamefont
  {Brady}},\ }\bibfield  {title} {\bibinfo {title} {Upstream swimming and
  taylor dispersion of active brownian particles},\ }\href
  {https://doi.org/10.1103/PhysRevFluids.5.073102} {\bibfield  {journal}
  {\bibinfo  {journal} {Physical Review Fluids}\ }\textbf {\bibinfo {volume}
  {5}},\ \bibinfo {pages} {073102} (\bibinfo {year} {2020})}\BibitemShut
  {NoStop}%
\bibitem [{\citenamefont {Ezhilan}\ and\ \citenamefont
  {Saintillan}(2015)}]{Ezhilan:2015}%
  \BibitemOpen
  \bibfield  {author} {\bibinfo {author} {\bibfnamefont {B.}~\bibnamefont
  {Ezhilan}}\ and\ \bibinfo {author} {\bibfnamefont {D.}~\bibnamefont
  {Saintillan}},\ }\bibfield  {title} {\bibinfo {title} {Transport of a dilute
  active suspension in pressure-driven channel flow},\ }\href
  {https://doi.org/10.1017/jfm.2015.372} {\bibfield  {journal} {\bibinfo
  {journal} {Journal of Fluid Mechanics}\ }\textbf {\bibinfo {volume} {777}},\
  \bibinfo {pages} {482–522} (\bibinfo {year} {2015})}\BibitemShut {NoStop}%
\bibitem [{\citenamefont {Marcos}\ \emph {et~al.}(2012)\citenamefont {Marcos},
  \citenamefont {Fu}, \citenamefont {Powers},\ and\ \citenamefont
  {Stocker}}]{Marcos:2012}%
  \BibitemOpen
  \bibfield  {author} {\bibinfo {author} {\bibnamefont {Marcos}}, \bibinfo
  {author} {\bibfnamefont {H.~C.}\ \bibnamefont {Fu}}, \bibinfo {author}
  {\bibfnamefont {T.~R.}\ \bibnamefont {Powers}},\ and\ \bibinfo {author}
  {\bibfnamefont {R.}~\bibnamefont {Stocker}},\ }\bibfield  {title} {\bibinfo
  {title} {Bacterial rheotaxis},\ }\href
  {https://doi.org/10.1073/pnas.1120955109} {\bibfield  {journal} {\bibinfo
  {journal} {Proceedings of the National Academy of Sciences}\ }\textbf
  {\bibinfo {volume} {109}},\ \bibinfo {pages} {4780} (\bibinfo {year}
  {2012})}\BibitemShut {NoStop}%
\bibitem [{\citenamefont {Lagoin}\ \emph {et~al.}(2025)\citenamefont {Lagoin},
  \citenamefont {Lacherez}, \citenamefont {Tournemire}, \citenamefont {Badr},
  \citenamefont {Amarouchene}, \citenamefont {Allard},\ and\ \citenamefont
  {Salez}}]{Lagoin:2025}%
  \BibitemOpen
  \bibfield  {author} {\bibinfo {author} {\bibfnamefont {M.}~\bibnamefont
  {Lagoin}}, \bibinfo {author} {\bibfnamefont {J.}~\bibnamefont {Lacherez}},
  \bibinfo {author} {\bibfnamefont {G.~d.}\ \bibnamefont {Tournemire}},
  \bibinfo {author} {\bibfnamefont {A.}~\bibnamefont {Badr}}, \bibinfo {author}
  {\bibfnamefont {Y.}~\bibnamefont {Amarouchene}}, \bibinfo {author}
  {\bibfnamefont {A.}~\bibnamefont {Allard}},\ and\ \bibinfo {author}
  {\bibfnamefont {T.}~\bibnamefont {Salez}},\ }\bibfield  {title} {\bibinfo
  {title} {Enhanced dispersion of active microswimmers in confined flows}\
  }\href {https://doi.org/10.48550/arXiv.2507.08369}
  {10.48550/arXiv.2507.08369} (\bibinfo {year} {2025}),\ \bibinfo {note}
  {arXiv:2507.08369 [cond-mat]}\BibitemShut {NoStop}%
\bibitem [{\citenamefont {Rusconi}\ \emph {et~al.}(2014)\citenamefont
  {Rusconi}, \citenamefont {Guasto},\ and\ \citenamefont
  {Stocker}}]{Rusconi:2014}%
  \BibitemOpen
  \bibfield  {author} {\bibinfo {author} {\bibfnamefont {R.}~\bibnamefont
  {Rusconi}}, \bibinfo {author} {\bibfnamefont {J.~S.}\ \bibnamefont
  {Guasto}},\ and\ \bibinfo {author} {\bibfnamefont {R.}~\bibnamefont
  {Stocker}},\ }\bibfield  {title} {\bibinfo {title} {Bacterial transport
  suppressed by fluid shear},\ }\href {https://doi.org/10.1038/nphys2883}
  {\bibfield  {journal} {\bibinfo  {journal} {Nature Physics}\ }\textbf
  {\bibinfo {volume} {10}},\ \bibinfo {pages} {212–217} (\bibinfo {year}
  {2014})}\BibitemShut {NoStop}%
\bibitem [{\citenamefont {Barry}\ \emph {et~al.}(2015)\citenamefont {Barry},
  \citenamefont {Rusconi}, \citenamefont {Guasto},\ and\ \citenamefont
  {Stocker}}]{Barry:2015}%
  \BibitemOpen
  \bibfield  {author} {\bibinfo {author} {\bibfnamefont {M.~T.}\ \bibnamefont
  {Barry}}, \bibinfo {author} {\bibfnamefont {R.}~\bibnamefont {Rusconi}},
  \bibinfo {author} {\bibfnamefont {J.~S.}\ \bibnamefont {Guasto}},\ and\
  \bibinfo {author} {\bibfnamefont {R.}~\bibnamefont {Stocker}},\ }\bibfield
  {title} {\bibinfo {title} {Shear-induced orientational dynamics and spatial
  heterogeneity in suspensions of motile phytoplankton},\ }\href
  {https://doi.org/10.1098/rsif.2015.0791} {\bibfield  {journal} {\bibinfo
  {journal} {Journal of The Royal Society Interface}\ }\textbf {\bibinfo
  {volume} {12}},\ \bibinfo {pages} {20150791} (\bibinfo {year}
  {2015})}\BibitemShut {NoStop}%
\bibitem [{\citenamefont {Vennamneni}\ \emph {et~al.}(2020)\citenamefont
  {Vennamneni}, \citenamefont {Nambiar},\ and\ \citenamefont
  {Subramanian}}]{Vennamneni:2020}%
  \BibitemOpen
  \bibfield  {author} {\bibinfo {author} {\bibfnamefont {L.}~\bibnamefont
  {Vennamneni}}, \bibinfo {author} {\bibfnamefont {S.}~\bibnamefont
  {Nambiar}},\ and\ \bibinfo {author} {\bibfnamefont {G.}~\bibnamefont
  {Subramanian}},\ }\bibfield  {title} {\bibinfo {title} {Shear-induced
  migration of microswimmers in pressure-driven channel flow},\ }\href
  {https://doi.org/10.1017/jfm.2020.118} {\bibfield  {journal} {\bibinfo
  {journal} {Journal of Fluid Mechanics}\ }\textbf {\bibinfo {volume} {890}},\
  \bibinfo {pages} {A15} (\bibinfo {year} {2020})}\BibitemShut {NoStop}%
\bibitem [{\citenamefont {Miño}\ \emph {et~al.}(2018)\citenamefont {Miño},
  \citenamefont {Baabour}, \citenamefont {Chertcoff}, \citenamefont {Gutkind},
  \citenamefont {Clément}, \citenamefont {Auradou},\ and\ \citenamefont
  {Ippolito}}]{Mino:2018}%
  \BibitemOpen
  \bibfield  {author} {\bibinfo {author} {\bibfnamefont {G.~L.}\ \bibnamefont
  {Miño}}, \bibinfo {author} {\bibfnamefont {M.}~\bibnamefont {Baabour}},
  \bibinfo {author} {\bibfnamefont {R.}~\bibnamefont {Chertcoff}}, \bibinfo
  {author} {\bibfnamefont {G.}~\bibnamefont {Gutkind}}, \bibinfo {author}
  {\bibfnamefont {E.}~\bibnamefont {Clément}}, \bibinfo {author}
  {\bibfnamefont {H.}~\bibnamefont {Auradou}},\ and\ \bibinfo {author}
  {\bibfnamefont {I.}~\bibnamefont {Ippolito}},\ }\bibfield  {title} {\bibinfo
  {title} {E. coli accumulation behind an obstacle},\ }\href
  {https://doi.org/10.4236/aim.2018.86030} {\bibfield  {journal} {\bibinfo
  {journal} {Advances in Microbiology}\ }\textbf {\bibinfo {volume} {08}},\
  \bibinfo {pages} {451–464} (\bibinfo {year} {2018})}\BibitemShut {NoStop}%
\bibitem [{\citenamefont {Secchi}\ \emph {et~al.}(2020)\citenamefont {Secchi},
  \citenamefont {Vitale}, \citenamefont {Miño}, \citenamefont {Kantsler},
  \citenamefont {Eberl}, \citenamefont {Rusconi},\ and\ \citenamefont
  {Stocker}}]{Secchi:2020}%
  \BibitemOpen
  \bibfield  {author} {\bibinfo {author} {\bibfnamefont {E.}~\bibnamefont
  {Secchi}}, \bibinfo {author} {\bibfnamefont {A.}~\bibnamefont {Vitale}},
  \bibinfo {author} {\bibfnamefont {G.~L.}\ \bibnamefont {Miño}}, \bibinfo
  {author} {\bibfnamefont {V.}~\bibnamefont {Kantsler}}, \bibinfo {author}
  {\bibfnamefont {L.}~\bibnamefont {Eberl}}, \bibinfo {author} {\bibfnamefont
  {R.}~\bibnamefont {Rusconi}},\ and\ \bibinfo {author} {\bibfnamefont
  {R.}~\bibnamefont {Stocker}},\ }\bibfield  {title} {\bibinfo {title} {The
  effect of flow on swimming bacteria controls the initial colonization of
  curved surfaces},\ }\href {https://doi.org/10.1038/s41467-020-16620-y}
  {\bibfield  {journal} {\bibinfo  {journal} {Nature Communications}\ }\textbf
  {\bibinfo {volume} {11}},\ \bibinfo {pages} {2851} (\bibinfo {year}
  {2020})}\BibitemShut {NoStop}%
\bibitem [{\citenamefont {Lee}\ \emph {et~al.}(2021)\citenamefont {Lee},
  \citenamefont {Lohrmann}, \citenamefont {Szuttor}, \citenamefont {Auradou},\
  and\ \citenamefont {Holm}}]{Lee:2021}%
  \BibitemOpen
  \bibfield  {author} {\bibinfo {author} {\bibfnamefont {M.}~\bibnamefont
  {Lee}}, \bibinfo {author} {\bibfnamefont {C.}~\bibnamefont {Lohrmann}},
  \bibinfo {author} {\bibfnamefont {K.}~\bibnamefont {Szuttor}}, \bibinfo
  {author} {\bibfnamefont {H.}~\bibnamefont {Auradou}},\ and\ \bibinfo {author}
  {\bibfnamefont {C.}~\bibnamefont {Holm}},\ }\bibfield  {title} {\bibinfo
  {title} {The influence of motility on bacterial accumulation in a microporous
  channel},\ }\href {https://doi.org/10.1039/D0SM01595D} {\bibfield  {journal}
  {\bibinfo  {journal} {Soft Matter}\ }\textbf {\bibinfo {volume} {17}},\
  \bibinfo {pages} {893–902} (\bibinfo {year} {2021})}\BibitemShut {NoStop}%
\bibitem [{\citenamefont {Yan}\ and\ \citenamefont {Brady}(2015)}]{Yan:2015}%
  \BibitemOpen
  \bibfield  {author} {\bibinfo {author} {\bibfnamefont {W.}~\bibnamefont
  {Yan}}\ and\ \bibinfo {author} {\bibfnamefont {J.~F.}\ \bibnamefont
  {Brady}},\ }\bibfield  {title} {\bibinfo {title} {The force on a boundary in
  active matter},\ }\href {https://doi.org/10.1017/jfm.2015.621} {\bibfield
  {journal} {\bibinfo  {journal} {Journal of Fluid Mechanics}\ }\textbf
  {\bibinfo {volume} {785}},\ \bibinfo {pages} {R1} (\bibinfo {year}
  {2015})}\BibitemShut {NoStop}%
\bibitem [{\citenamefont {Dentz}\ \emph {et~al.}(2022)\citenamefont {Dentz},
  \citenamefont {Creppy}, \citenamefont {Douarche}, \citenamefont {Clément},\
  and\ \citenamefont {Auradou}}]{Dentz:2022}%
  \BibitemOpen
  \bibfield  {author} {\bibinfo {author} {\bibfnamefont {M.}~\bibnamefont
  {Dentz}}, \bibinfo {author} {\bibfnamefont {A.}~\bibnamefont {Creppy}},
  \bibinfo {author} {\bibfnamefont {C.}~\bibnamefont {Douarche}}, \bibinfo
  {author} {\bibfnamefont {E.}~\bibnamefont {Clément}},\ and\ \bibinfo
  {author} {\bibfnamefont {H.}~\bibnamefont {Auradou}},\ }\bibfield  {title}
  {\bibinfo {title} {Dispersion of motile bacteria in a porous medium},\ }\href
  {https://doi.org/10.1017/jfm.2022.596} {\bibfield  {journal} {\bibinfo
  {journal} {Journal of Fluid Mechanics}\ }\textbf {\bibinfo {volume} {946}},\
  \bibinfo {pages} {A33} (\bibinfo {year} {2022})}\BibitemShut {NoStop}%
\bibitem [{\citenamefont {Altshuler}\ \emph {et~al.}(2013)\citenamefont
  {Altshuler}, \citenamefont {Miño}, \citenamefont {Pérez-Penichet},
  \citenamefont {Río}, \citenamefont {Lindner}, \citenamefont {Rousselet},\
  and\ \citenamefont {Clément}}]{Altshuler:2013}%
  \BibitemOpen
  \bibfield  {author} {\bibinfo {author} {\bibfnamefont {E.}~\bibnamefont
  {Altshuler}}, \bibinfo {author} {\bibfnamefont {G.}~\bibnamefont {Miño}},
  \bibinfo {author} {\bibfnamefont {C.}~\bibnamefont {Pérez-Penichet}},
  \bibinfo {author} {\bibfnamefont {L.~d.}\ \bibnamefont {Río}}, \bibinfo
  {author} {\bibfnamefont {A.}~\bibnamefont {Lindner}}, \bibinfo {author}
  {\bibfnamefont {A.}~\bibnamefont {Rousselet}},\ and\ \bibinfo {author}
  {\bibfnamefont {E.}~\bibnamefont {Clément}},\ }\bibfield  {title} {\bibinfo
  {title} {Flow-controlled densification and anomalous dispersion of e. coli
  through a constriction},\ }\href {https://doi.org/10.1039/C2SM26460A}
  {\bibfield  {journal} {\bibinfo  {journal} {Soft Matter}\ }\textbf {\bibinfo
  {volume} {9}},\ \bibinfo {pages} {1864–1870} (\bibinfo {year}
  {2013})}\BibitemShut {NoStop}%
\bibitem [{\citenamefont {Waisbord}\ \emph {et~al.}(2021)\citenamefont
  {Waisbord}, \citenamefont {Dehkharghani},\ and\ \citenamefont
  {Guasto}}]{waisbord2021fluidic}%
  \BibitemOpen
  \bibfield  {author} {\bibinfo {author} {\bibfnamefont {N.}~\bibnamefont
  {Waisbord}}, \bibinfo {author} {\bibfnamefont {A.}~\bibnamefont
  {Dehkharghani}},\ and\ \bibinfo {author} {\bibfnamefont {J.~S.}\ \bibnamefont
  {Guasto}},\ }\bibfield  {title} {\bibinfo {title} {Fluidic bacterial diodes
  rectify magnetotactic cell motility in porous environments},\ }\href
  {https://doi.org/10.1038/s41467-021-26235-6} {\bibfield  {journal} {\bibinfo
  {journal} {Nature Communications}\ }\textbf {\bibinfo {volume} {12}},\
  \bibinfo {pages} {5949} (\bibinfo {year} {2021})}\BibitemShut {NoStop}%
\bibitem [{\citenamefont {Bhattacharjee}\ and\ \citenamefont
  {Datta}(2019)}]{Bhattacharjee:2019}%
  \BibitemOpen
  \bibfield  {author} {\bibinfo {author} {\bibfnamefont {T.}~\bibnamefont
  {Bhattacharjee}}\ and\ \bibinfo {author} {\bibfnamefont {S.~S.}\ \bibnamefont
  {Datta}},\ }\bibfield  {title} {\bibinfo {title} {Bacterial hopping and
  trapping in porous media},\ }\href
  {https://doi.org/10.1038/s41467-019-10115-1} {\bibfield  {journal} {\bibinfo
  {journal} {Nature Communications}\ }\textbf {\bibinfo {volume} {10}},\
  \bibinfo {pages} {1} (\bibinfo {year} {2019})}\BibitemShut {NoStop}%
\bibitem [{\citenamefont {Bertrand}\ \emph {et~al.}(2018)\citenamefont
  {Bertrand}, \citenamefont {Zhao}, \citenamefont {B\'enichou}, \citenamefont
  {Tailleur},\ and\ \citenamefont {Voituriez}}]{Bertrand:2018}%
  \BibitemOpen
  \bibfield  {author} {\bibinfo {author} {\bibfnamefont {T.}~\bibnamefont
  {Bertrand}}, \bibinfo {author} {\bibfnamefont {Y.}~\bibnamefont {Zhao}},
  \bibinfo {author} {\bibfnamefont {O.}~\bibnamefont {B\'enichou}}, \bibinfo
  {author} {\bibfnamefont {J.}~\bibnamefont {Tailleur}},\ and\ \bibinfo
  {author} {\bibfnamefont {R.}~\bibnamefont {Voituriez}},\ }\bibfield  {title}
  {\bibinfo {title} {Optimized diffusion of run-and-tumble particles in crowded
  environments},\ }\href {https://doi.org/10.1103/PhysRevLett.120.198103}
  {\bibfield  {journal} {\bibinfo  {journal} {Physical Review Letters}\
  }\textbf {\bibinfo {volume} {120}},\ \bibinfo {pages} {198103} (\bibinfo
  {year} {2018})}\BibitemShut {NoStop}%
\bibitem [{\citenamefont {Licata}\ \emph {et~al.}(2016)\citenamefont {Licata},
  \citenamefont {Mohari}, \citenamefont {Fuqua},\ and\ \citenamefont
  {Setayeshgar}}]{Licata:2016}%
  \BibitemOpen
  \bibfield  {author} {\bibinfo {author} {\bibfnamefont {N.~A.}\ \bibnamefont
  {Licata}}, \bibinfo {author} {\bibfnamefont {B.}~\bibnamefont {Mohari}},
  \bibinfo {author} {\bibfnamefont {C.}~\bibnamefont {Fuqua}},\ and\ \bibinfo
  {author} {\bibfnamefont {S.}~\bibnamefont {Setayeshgar}},\ }\bibfield
  {title} {\bibinfo {title} {Diffusion of bacterial cells in porous media},\
  }\href {https://doi.org/10.1016/j.bpj.2015.09.035} {\bibfield  {journal}
  {\bibinfo  {journal} {Biophysical Journal}\ }\textbf {\bibinfo {volume}
  {110}},\ \bibinfo {pages} {247} (\bibinfo {year} {2016})}\BibitemShut
  {NoStop}%
\bibitem [{\citenamefont {Volpe}\ and\ \citenamefont
  {Volpe}(2017)}]{Volpe:2017}%
  \BibitemOpen
  \bibfield  {author} {\bibinfo {author} {\bibfnamefont {G.}~\bibnamefont
  {Volpe}}\ and\ \bibinfo {author} {\bibfnamefont {G.}~\bibnamefont {Volpe}},\
  }\bibfield  {title} {\bibinfo {title} {The topography of the environment
  alters the optimal search strategy for active particles},\ }\href
  {https://doi.org/10.1073/pnas.1711371114} {\bibfield  {journal} {\bibinfo
  {journal} {Proceedings of the National Academy of Sciences of the United
  States of America}\ }\textbf {\bibinfo {volume} {114}},\ \bibinfo {pages}
  {11350} (\bibinfo {year} {2017})}\BibitemShut {NoStop}%
\bibitem [{\citenamefont {Kurzthaler}\ \emph {et~al.}(2021)\citenamefont
  {Kurzthaler}, \citenamefont {Mandal}, \citenamefont {Bhattacharjee},
  \citenamefont {L{\"o}wen}, \citenamefont {Datta},\ and\ \citenamefont
  {Stone}}]{Kurzthaler:2021}%
  \BibitemOpen
  \bibfield  {author} {\bibinfo {author} {\bibfnamefont {C.}~\bibnamefont
  {Kurzthaler}}, \bibinfo {author} {\bibfnamefont {S.}~\bibnamefont {Mandal}},
  \bibinfo {author} {\bibfnamefont {T.}~\bibnamefont {Bhattacharjee}}, \bibinfo
  {author} {\bibfnamefont {H.}~\bibnamefont {L{\"o}wen}}, \bibinfo {author}
  {\bibfnamefont {S.~S.}\ \bibnamefont {Datta}},\ and\ \bibinfo {author}
  {\bibfnamefont {H.~A.}\ \bibnamefont {Stone}},\ }\bibfield  {title} {\bibinfo
  {title} {A geometric criterion for the optimal spreading of active polymers
  in porous media},\ }\href {https://doi.org/10.1038/s41467-021-26942-0}
  {\bibfield  {journal} {\bibinfo  {journal} {Nature communications}\ }\textbf
  {\bibinfo {volume} {12}},\ \bibinfo {pages} {7088} (\bibinfo {year}
  {2021})}\BibitemShut {NoStop}%
\bibitem [{\citenamefont {Reichhardt}\ and\ \citenamefont
  {Olson~Reichhardt}(2014)}]{Reichhardt:2014}%
  \BibitemOpen
  \bibfield  {author} {\bibinfo {author} {\bibfnamefont {C.}~\bibnamefont
  {Reichhardt}}\ and\ \bibinfo {author} {\bibfnamefont {C.~J.}\ \bibnamefont
  {Olson~Reichhardt}},\ }\bibfield  {title} {\bibinfo {title} {Active matter
  transport and jamming on disordered landscapes},\ }\href
  {https://doi.org/10.1103/PhysRevE.90.012701} {\bibfield  {journal} {\bibinfo
  {journal} {Physical Review E}\ }\textbf {\bibinfo {volume} {90}},\ \bibinfo
  {pages} {012701} (\bibinfo {year} {2014})}\BibitemShut {NoStop}%
\bibitem [{\citenamefont {Residori}\ \emph {et~al.}(2025)\citenamefont
  {Residori}, \citenamefont {Mandal}, \citenamefont {Voigt},\ and\
  \citenamefont {Kurzthaler}}]{Residori:2025}%
  \BibitemOpen
  \bibfield  {author} {\bibinfo {author} {\bibfnamefont {M.}~\bibnamefont
  {Residori}}, \bibinfo {author} {\bibfnamefont {S.}~\bibnamefont {Mandal}},
  \bibinfo {author} {\bibfnamefont {A.}~\bibnamefont {Voigt}},\ and\ \bibinfo
  {author} {\bibfnamefont {C.}~\bibnamefont {Kurzthaler}},\ }\bibfield  {title}
  {\bibinfo {title} {Flow through porous media at the percolation transition},\
  }\href {https://doi.org/10.1103/PhysRevResearch.7.L012032} {\bibfield
  {journal} {\bibinfo  {journal} {Physical Review Research}\ }\textbf {\bibinfo
  {volume} {7}},\ \bibinfo {pages} {L012032} (\bibinfo {year}
  {2025})}\BibitemShut {NoStop}%
\bibitem [{\citenamefont {Aufrecht}\ \emph {et~al.}(2019)\citenamefont
  {Aufrecht}, \citenamefont {Fowlkes}, \citenamefont {Bible}, \citenamefont
  {Morrell-Falvey}, \citenamefont {Doktycz},\ and\ \citenamefont
  {Retterer}}]{Aufrecht:2019}%
  \BibitemOpen
  \bibfield  {author} {\bibinfo {author} {\bibfnamefont {J.~A.}\ \bibnamefont
  {Aufrecht}}, \bibinfo {author} {\bibfnamefont {J.~D.}\ \bibnamefont
  {Fowlkes}}, \bibinfo {author} {\bibfnamefont {A.~N.}\ \bibnamefont {Bible}},
  \bibinfo {author} {\bibfnamefont {J.}~\bibnamefont {Morrell-Falvey}},
  \bibinfo {author} {\bibfnamefont {M.~J.}\ \bibnamefont {Doktycz}},\ and\
  \bibinfo {author} {\bibfnamefont {S.~T.}\ \bibnamefont {Retterer}},\
  }\bibfield  {title} {\bibinfo {title} {Pore-scale hydrodynamics influence the
  spatial evolution of bacterial biofilms in a microfluidic porous network},\
  }\href {https://doi.org/10.1371/journal.pone.0218316} {\bibfield  {journal}
  {\bibinfo  {journal} {PLOS ONE}\ }\textbf {\bibinfo {volume} {14}},\ \bibinfo
  {pages} {e0218316} (\bibinfo {year} {2019})}\BibitemShut {NoStop}%
\bibitem [{\citenamefont {Kurz}\ \emph {et~al.}(2022)\citenamefont {Kurz},
  \citenamefont {Secchi}, \citenamefont {Carrillo}, \citenamefont {Bourg},
  \citenamefont {Stocker},\ and\ \citenamefont {Jimenez-Martinez}}]{Kurz:2022}%
  \BibitemOpen
  \bibfield  {author} {\bibinfo {author} {\bibfnamefont {D.~L.}\ \bibnamefont
  {Kurz}}, \bibinfo {author} {\bibfnamefont {E.}~\bibnamefont {Secchi}},
  \bibinfo {author} {\bibfnamefont {F.~J.}\ \bibnamefont {Carrillo}}, \bibinfo
  {author} {\bibfnamefont {I.~C.}\ \bibnamefont {Bourg}}, \bibinfo {author}
  {\bibfnamefont {R.}~\bibnamefont {Stocker}},\ and\ \bibinfo {author}
  {\bibfnamefont {J.}~\bibnamefont {Jimenez-Martinez}},\ }\bibfield  {title}
  {\bibinfo {title} {Competition between growth and shear stress drives
  intermittency in preferential flow paths in porous medium biofilms},\ }\href
  {https://doi.org/10.1073/pnas.2122202119} {\bibfield  {journal} {\bibinfo
  {journal} {Proceedings of the National Academy of Sciences}\ }\textbf
  {\bibinfo {volume} {119}},\ \bibinfo {pages} {e2122202119} (\bibinfo {year}
  {2022})}\BibitemShut {NoStop}%
\bibitem [{\citenamefont {Kurz}\ \emph {et~al.}(2023)\citenamefont {Kurz},
  \citenamefont {Secchi}, \citenamefont {Stocker},\ and\ \citenamefont
  {Jimenez-Martinez}}]{Kurz:2023}%
  \BibitemOpen
  \bibfield  {author} {\bibinfo {author} {\bibfnamefont {D.~L.}\ \bibnamefont
  {Kurz}}, \bibinfo {author} {\bibfnamefont {E.}~\bibnamefont {Secchi}},
  \bibinfo {author} {\bibfnamefont {R.}~\bibnamefont {Stocker}},\ and\ \bibinfo
  {author} {\bibfnamefont {J.}~\bibnamefont {Jimenez-Martinez}},\ }\bibfield
  {title} {\bibinfo {title} {Morphogenesis of biofilms in porous media and
  control on hydrodynamics},\ }\href {https://doi.org/10.1021/acs.est.2c08890}
  {\bibfield  {journal} {\bibinfo  {journal} {Environmental Science \&
  Technology}\ }\textbf {\bibinfo {volume} {57}},\ \bibinfo {pages}
  {5666–5677} (\bibinfo {year} {2023})}\BibitemShut {NoStop}%
\bibitem [{\citenamefont {Howse}\ \emph {et~al.}(2007)\citenamefont {Howse},
  \citenamefont {Jones}, \citenamefont {Ryan}, \citenamefont {Gough},
  \citenamefont {Vafabakhsh},\ and\ \citenamefont {Golestanian}}]{Howse:2007}%
  \BibitemOpen
  \bibfield  {author} {\bibinfo {author} {\bibfnamefont {J.~R.}\ \bibnamefont
  {Howse}}, \bibinfo {author} {\bibfnamefont {R.~A.~L.}\ \bibnamefont {Jones}},
  \bibinfo {author} {\bibfnamefont {A.~J.}\ \bibnamefont {Ryan}}, \bibinfo
  {author} {\bibfnamefont {T.}~\bibnamefont {Gough}}, \bibinfo {author}
  {\bibfnamefont {R.}~\bibnamefont {Vafabakhsh}},\ and\ \bibinfo {author}
  {\bibfnamefont {R.}~\bibnamefont {Golestanian}},\ }\bibfield  {title}
  {\bibinfo {title} {Self-motile colloidal particles: From directed propulsion
  to random walk},\ }\href {https://doi.org/10.1103/PhysRevLett.99.048102}
  {\bibfield  {journal} {\bibinfo  {journal} {Physical Review Letters}\
  }\textbf {\bibinfo {volume} {99}},\ \bibinfo {pages} {048102} (\bibinfo
  {year} {2007})}\BibitemShut {NoStop}%
\bibitem [{\citenamefont {Romanczuk}\ \emph {et~al.}(2012)\citenamefont
  {Romanczuk}, \citenamefont {Bär}, \citenamefont {Ebeling}, \citenamefont
  {Lindner},\ and\ \citenamefont {Schimansky-Geier}}]{Romanczuk:2012}%
  \BibitemOpen
  \bibfield  {author} {\bibinfo {author} {\bibfnamefont {P.}~\bibnamefont
  {Romanczuk}}, \bibinfo {author} {\bibfnamefont {M.}~\bibnamefont {Bär}},
  \bibinfo {author} {\bibfnamefont {W.}~\bibnamefont {Ebeling}}, \bibinfo
  {author} {\bibfnamefont {B.}~\bibnamefont {Lindner}},\ and\ \bibinfo {author}
  {\bibfnamefont {L.}~\bibnamefont {Schimansky-Geier}},\ }\bibfield  {title}
  {\bibinfo {title} {Active brownian particles},\ }\href
  {https://doi.org/10.1140/epjst/e2012-01529-y} {\bibfield  {journal} {\bibinfo
   {journal} {The European Physical Journal Special Topics}\ }\textbf {\bibinfo
  {volume} {202}},\ \bibinfo {pages} {1–162} (\bibinfo {year}
  {2012})}\BibitemShut {NoStop}%
\bibitem [{\citenamefont {Kurzthaler}\ \emph {et~al.}(2016)\citenamefont
  {Kurzthaler}, \citenamefont {Leitmann},\ and\ \citenamefont
  {Franosch}}]{Kurzthaler:2016}%
  \BibitemOpen
  \bibfield  {author} {\bibinfo {author} {\bibfnamefont {C.}~\bibnamefont
  {Kurzthaler}}, \bibinfo {author} {\bibfnamefont {S.}~\bibnamefont
  {Leitmann}},\ and\ \bibinfo {author} {\bibfnamefont {T.}~\bibnamefont
  {Franosch}},\ }\bibfield  {title} {\bibinfo {title} {Intermediate scattering
  function of an anisotropic active brownian particle},\ }\href
  {https://doi.org/10.1038/srep36702} {\bibfield  {journal} {\bibinfo
  {journal} {Scientific Reports}\ }\textbf {\bibinfo {volume} {6}},\ \bibinfo
  {pages} {36702} (\bibinfo {year} {2016})}\BibitemShut {NoStop}%
\bibitem [{\citenamefont {Leal}(2007)}]{leal2007advanced}%
  \BibitemOpen
  \bibfield  {author} {\bibinfo {author} {\bibfnamefont {L.~G.}\ \bibnamefont
  {Leal}},\ }\href {https://doi.org/10.1017/CBO9780511800245} {\emph {\bibinfo
  {title} {Advanced {T}ransport {P}henomena}}},\ Vol.~\bibinfo {volume} {7}\
  (\bibinfo  {publisher} {Cambridge University Press},\ \bibinfo {year}
  {2007})\BibitemShut {NoStop}%
\bibitem [{\citenamefont {Zeitz}\ \emph {et~al.}(2017)\citenamefont {Zeitz},
  \citenamefont {Wolff},\ and\ \citenamefont {Stark}}]{Zeitz:2017}%
  \BibitemOpen
  \bibfield  {author} {\bibinfo {author} {\bibfnamefont {M.}~\bibnamefont
  {Zeitz}}, \bibinfo {author} {\bibfnamefont {K.}~\bibnamefont {Wolff}},\ and\
  \bibinfo {author} {\bibfnamefont {H.}~\bibnamefont {Stark}},\ }\bibfield
  {title} {\bibinfo {title} {Active brownian particles moving in a random
  lorentz gas},\ }\href {https://doi.org/10.1140/epje/i2017-11510-0} {\bibfield
   {journal} {\bibinfo  {journal} {The European Physical Journal E}\ }\textbf
  {\bibinfo {volume} {40}},\ \bibinfo {pages} {23} (\bibinfo {year}
  {2017})}\BibitemShut {NoStop}%
\bibitem [{\citenamefont {Torquato}(2002)}]{torquato_random_2002}%
  \BibitemOpen
  \bibfield  {author} {\bibinfo {author} {\bibfnamefont {S.}~\bibnamefont
  {Torquato}},\ }\href {https://doi.org/10.1007/978-1-4757-6355-3} {\emph
  {\bibinfo {title} {Random {Heterogeneous} {Materials}}}},\ edited by\
  \bibinfo {editor} {\bibfnamefont {S.~S.}\ \bibnamefont {Antman}}, \bibinfo
  {editor} {\bibfnamefont {L.}~\bibnamefont {Sirovich}}, \bibinfo {editor}
  {\bibfnamefont {J.~E.}\ \bibnamefont {Marsden}},\ and\ \bibinfo {editor}
  {\bibfnamefont {S.}~\bibnamefont {Wiggins}},\ \bibinfo {series}
  {Interdisciplinary {Applied} {Mathematics}}, Vol.~\bibinfo {volume} {16}\
  (\bibinfo  {publisher} {Springer New York},\ \bibinfo {address} {New York,
  NY},\ \bibinfo {year} {2002})\BibitemShut {NoStop}%
\bibitem [{\citenamefont {Bouchaud}\ and\ \citenamefont
  {Georges}(1990)}]{bouchaud_anomalous_1990}%
  \BibitemOpen
  \bibfield  {author} {\bibinfo {author} {\bibfnamefont {J.-P.}\ \bibnamefont
  {Bouchaud}}\ and\ \bibinfo {author} {\bibfnamefont {A.}~\bibnamefont
  {Georges}},\ }\bibfield  {title} {\bibinfo {title} {Anomalous diffusion in
  disordered media: {Statistical} mechanisms, models and physical
  applications},\ }\href {https://doi.org/10.1016/0370-1573(90)90099-N}
  {\bibfield  {journal} {\bibinfo  {journal} {Physics Reports}\ }\textbf
  {\bibinfo {volume} {195}},\ \bibinfo {pages} {127} (\bibinfo {year}
  {1990})}\BibitemShut {NoStop}%
\bibitem [{\citenamefont {Baouche}\ \emph
  {et~al.}(2025{\natexlab{a}})\citenamefont {Baouche}, \citenamefont {Le~Goff},
  \citenamefont {Kurzthaler},\ and\ \citenamefont {Franosch}}]{Baouche:2025}%
  \BibitemOpen
  \bibfield  {author} {\bibinfo {author} {\bibfnamefont {Y.}~\bibnamefont
  {Baouche}}, \bibinfo {author} {\bibfnamefont {M.}~\bibnamefont {Le~Goff}},
  \bibinfo {author} {\bibfnamefont {C.}~\bibnamefont {Kurzthaler}},\ and\
  \bibinfo {author} {\bibfnamefont {T.}~\bibnamefont {Franosch}},\ }\bibfield
  {title} {\bibinfo {title} {First-passage-time statistics of active brownian
  particles: A perturbative approach},\ }\href
  {https://doi.org/10.1103/PhysRevE.111.054113} {\bibfield  {journal} {\bibinfo
   {journal} {Physical Review E}\ }\textbf {\bibinfo {volume} {111}},\ \bibinfo
  {pages} {054113} (\bibinfo {year} {2025}{\natexlab{a}})}\BibitemShut
  {NoStop}%
\bibitem [{\citenamefont {Baouche}\ \emph
  {et~al.}(2025{\natexlab{b}})\citenamefont {Baouche}, \citenamefont {Goff},
  \citenamefont {Franosch},\ and\ \citenamefont
  {Kurzthaler}}]{Baouche:2025:arxiv}%
  \BibitemOpen
  \bibfield  {author} {\bibinfo {author} {\bibfnamefont {Y.}~\bibnamefont
  {Baouche}}, \bibinfo {author} {\bibfnamefont {M.~L.}\ \bibnamefont {Goff}},
  \bibinfo {author} {\bibfnamefont {T.}~\bibnamefont {Franosch}},\ and\
  \bibinfo {author} {\bibfnamefont {C.}~\bibnamefont {Kurzthaler}},\ }\bibfield
   {title} {\bibinfo {title} {Hydrodynamic attraction and hindered diffusion
  govern first-passage times of swimming microorganisms},\ }\href
  {https://doi.org/10.48550/arXiv.2509.14765} {\bibfield  {journal} {\bibinfo
  {journal} {arXiv}\ ,\ \bibinfo {pages} {2509.14765}} (\bibinfo {year}
  {2025}{\natexlab{b}})}\BibitemShut {NoStop}%
\bibitem [{\citenamefont {Creppy}\ \emph {et~al.}(2019)\citenamefont {Creppy},
  \citenamefont {Clément}, \citenamefont {Douarche}, \citenamefont
  {D'Angelo},\ and\ \citenamefont {Auradou}}]{creppy_effect_2019}%
  \BibitemOpen
  \bibfield  {author} {\bibinfo {author} {\bibfnamefont {A.}~\bibnamefont
  {Creppy}}, \bibinfo {author} {\bibfnamefont {E.}~\bibnamefont {Clément}},
  \bibinfo {author} {\bibfnamefont {C.}~\bibnamefont {Douarche}}, \bibinfo
  {author} {\bibfnamefont {M.~V.}\ \bibnamefont {D'Angelo}},\ and\ \bibinfo
  {author} {\bibfnamefont {H.}~\bibnamefont {Auradou}},\ }\bibfield  {title}
  {\bibinfo {title} {Effect of motility on the transport of bacteria
  populations through a porous medium},\ }\href
  {https://doi.org/10.1103/PhysRevFluids.4.013102} {\bibfield  {journal}
  {\bibinfo  {journal} {Physical Review Fluids}\ }\textbf {\bibinfo {volume}
  {4}},\ \bibinfo {pages} {013102} (\bibinfo {year} {2019})}\BibitemShut
  {NoStop}%
\bibitem [{\citenamefont {De~Anna}\ \emph {et~al.}(2021)\citenamefont
  {De~Anna}, \citenamefont {Pahlavan}, \citenamefont {Yawata}, \citenamefont
  {Stocker},\ and\ \citenamefont {Juanes}}]{DeAnna:2021}%
  \BibitemOpen
  \bibfield  {author} {\bibinfo {author} {\bibfnamefont {P.}~\bibnamefont
  {De~Anna}}, \bibinfo {author} {\bibfnamefont {A.~A.}\ \bibnamefont
  {Pahlavan}}, \bibinfo {author} {\bibfnamefont {Y.}~\bibnamefont {Yawata}},
  \bibinfo {author} {\bibfnamefont {R.}~\bibnamefont {Stocker}},\ and\ \bibinfo
  {author} {\bibfnamefont {R.}~\bibnamefont {Juanes}},\ }\bibfield  {title}
  {\bibinfo {title} {Chemotaxis under flow disorder shapes microbial dispersion
  in porous media},\ }\href {https://doi.org/10.1038/s41567-020-1002-x}
  {\bibfield  {journal} {\bibinfo  {journal} {Nature Physics}\ }\textbf
  {\bibinfo {volume} {17}},\ \bibinfo {pages} {68–73} (\bibinfo {year}
  {2021})}\BibitemShut {NoStop}%
\bibitem [{\citenamefont {Grognot}\ \emph {et~al.}(2023)\citenamefont
  {Grognot}, \citenamefont {Nam}, \citenamefont {Elson},\ and\ \citenamefont
  {Taute}}]{Grogno:2023}%
  \BibitemOpen
  \bibfield  {author} {\bibinfo {author} {\bibfnamefont {M.}~\bibnamefont
  {Grognot}}, \bibinfo {author} {\bibfnamefont {J.~W.}\ \bibnamefont {Nam}},
  \bibinfo {author} {\bibfnamefont {L.~E.}\ \bibnamefont {Elson}},\ and\
  \bibinfo {author} {\bibfnamefont {K.~M.}\ \bibnamefont {Taute}},\ }\bibfield
  {title} {\bibinfo {title} {Physiological adaptation in flagellar architecture
  improves vibrio alginolyticus chemotaxis in complex environments},\ }\href
  {https://doi.org/10.1073/pnas.2301873120} {\bibfield  {journal} {\bibinfo
  {journal} {Proceedings of the National Academy of Sciences}\ }\textbf
  {\bibinfo {volume} {120}},\ \bibinfo {pages} {e2301873120} (\bibinfo {year}
  {2023})}\BibitemShut {NoStop}%
\bibitem [{\citenamefont {Keogh}\ \emph {et~al.}(2024)\citenamefont {Keogh},
  \citenamefont {Kozhukhov}, \citenamefont {Thijssen},\ and\ \citenamefont
  {Shendruk}}]{Keogh:2024}%
  \BibitemOpen
  \bibfield  {author} {\bibinfo {author} {\bibfnamefont {R.~R.}\ \bibnamefont
  {Keogh}}, \bibinfo {author} {\bibfnamefont {T.}~\bibnamefont {Kozhukhov}},
  \bibinfo {author} {\bibfnamefont {K.}~\bibnamefont {Thijssen}},\ and\
  \bibinfo {author} {\bibfnamefont {T.~N.}\ \bibnamefont {Shendruk}},\
  }\bibfield  {title} {\bibinfo {title} {Active {D}arcy's law},\ }\href
  {https://doi.org/10.1103/PhysRevLett.132.188301} {\bibfield  {journal}
  {\bibinfo  {journal} {Physical Review Letters}\ }\textbf {\bibinfo {volume}
  {132}},\ \bibinfo {pages} {188301} (\bibinfo {year} {2024})}\BibitemShut
  {NoStop}%
\bibitem [{\citenamefont {Residori}\ \emph {et~al.}(2023)\citenamefont
  {Residori}, \citenamefont {Praetorius}, \citenamefont {de~Anna},\ and\
  \citenamefont {Voigt}}]{Residori:2023}%
  \BibitemOpen
  \bibfield  {author} {\bibinfo {author} {\bibfnamefont {M.}~\bibnamefont
  {Residori}}, \bibinfo {author} {\bibfnamefont {S.}~\bibnamefont
  {Praetorius}}, \bibinfo {author} {\bibfnamefont {P.}~\bibnamefont
  {de~Anna}},\ and\ \bibinfo {author} {\bibfnamefont {A.}~\bibnamefont
  {Voigt}},\ }\bibfield  {title} {\bibinfo {title} {Influence of finite-size
  particles on fluid velocity and transport through porous media},\ }\href
  {https://doi.org/10.1103/PhysRevFluids.8.074501} {\bibfield  {journal}
  {\bibinfo  {journal} {Physical Review Fluids}\ }\textbf {\bibinfo {volume}
  {8}},\ \bibinfo {pages} {074501} (\bibinfo {year} {2023})}\BibitemShut
  {NoStop}%
\bibitem [{\citenamefont {Elam}\ \emph {et~al.}(1984)\citenamefont {Elam},
  \citenamefont {Kerstein},\ and\ \citenamefont {Rehr}}]{elam_1984}%
  \BibitemOpen
  \bibfield  {author} {\bibinfo {author} {\bibfnamefont {W.~T.}\ \bibnamefont
  {Elam}}, \bibinfo {author} {\bibfnamefont {A.~R.}\ \bibnamefont {Kerstein}},\
  and\ \bibinfo {author} {\bibfnamefont {J.~J.}\ \bibnamefont {Rehr}},\
  }\bibfield  {title} {\bibinfo {title} {Critical properties of the void
  percolation problem for spheres},\ }\href
  {https://doi.org/10.1103/PhysRevLett.52.1516} {\bibfield  {journal} {\bibinfo
   {journal} {Physical Review Letters}\ }\textbf {\bibinfo {volume} {52}},\
  \bibinfo {pages} {1516} (\bibinfo {year} {1984})}\BibitemShut {NoStop}%
\end{thebibliography}%

\end{document}